\documentclass[prl,twocolumn, noshowpacs, preprintnumbers, amsmath,amssymb,nopreprintnumbers]{revtex4-2}

\usepackage{dsfont}
\usepackage{graphicx}
\usepackage{lineno}
\usepackage{dcolumn}
\usepackage{bm}
\usepackage{appendix}
 \usepackage{verbatim}

\usepackage{color,xcolor}

\usepackage{amsmath}
\usepackage{enumerate}
\usepackage[normalem]{ulem}
\makeatletter
  \let\@font@info\@gobble
  \let\@font@warning\@gobble
\makeatother

\begin{document}
\title{Exponential Tails and Asymmetry  Relations for the Spread of Biased Random Walks
}

\author{Stanislav Burov$^1$}
\email{stasbur@gmail.com}
\author{Wanli Wang$^2$}
\email{wanliiwang@163.com}
\author{Eli Barkai$^{1,3}$}
\affiliation{$^1$Physics Department,  Bar-Ilan University, Ramat Gan 5290002,
Israel.\\
$^2$Department of Applied Mathematics, Zhejiang University of Technology, Hangzhou 310023, China.
\\
$^3$Institute of Nanotechnology and Advanced Materials, Bar-Ilan University, Ramat-Gan 5290002, Israel.}

\pacs{PACS}

\begin{abstract}
Exponential, and not Gaussian, decay of  probability density functions was studied by Laplace in the context of his analysis of errors. Such Laplace  propagators for the diffusive motion of single particles in disordered media were recently observed in numerous experimental systems. 
What will happen to this  universality 
when an external driving force is applied?  
Using the ubiquitous  continuous time random walk with bias, and the  Crooks relation in conjunction with large deviations theory, we derive  two properties of  the positional probability density function $P_F(x,t)$ that hold for a wide spectrum of random walk models: 
(I) Universal asymmetric exponential decay of $P_F(X,t)$ for large $|X|$, and (II) Existence of a time transformation that for large $|X|$ allows to express $P_F(X,t)$ in terms of the propagator of the unbiased process (measured at a shorter time). 
These findings allow us to establish how the symmetric exponential-like tails,  measured in many unbiased processes, will transform into asymmetric Laplace tails when an external force is applied. 
 \end{abstract}

\maketitle

When a weak external force perturbs a thermal system, the consequences are the focus of  linear response theory \cite{kubotoda,Bouchaud1990Anomalous}.
Quantification of the response is usually achieved via averages. 
Probably the most famous example is the celebrated Einstein relation for a Brownian particle under the influence of external force $F$ where the mobility is $\mu=D /k_B T$, hence $\langle X \rangle =\frac{D}{k_B T}Ft$, i.e., 
 the average position $\langle X\rangle$, 
is linear in terms of the external force,  time $t$, diffusion coefficient $D$ and  inverse of the temperature $1/T$.
If we consider a  Brownian particle  driven by  external force $F$, the positional probability density function (PDF) of $X$ is provided by the Gaussian
$P_F(X,t)=\exp\left[-(X-\frac{DFt}{k_B T})^2/4Dt\right]/\sqrt{4\pi Dt}$ (when starting from $X=0$). 
It is easily spotted that
\begin{equation}
    \label{eq:genassrel}
    R_{FF}(X)=\ln\left[P_F(X,t)\Big/P_F(-X,t)\right]
    ={F X}\Big/{k_B T}
\end{equation}
independent of $D$.
 Eq.~\eqref{eq:genassrel} describes the left-right asymmetry of the propagator and was obtained here for a Gaussian process, but it is much more general.
 Since $FX$ is the work performed by the external force $F$, Eq.~\eqref{eq:genassrel} is a manifestation of the Crooks fluctuation theorem~\cite{Crooks199,Crooks200} that was verified experimentally~\cite{bustamante2005} and has been the focus of many theoretical studies considering Markovian and non-Markovian dynamics~\cite{seifert2007,esposito2008,bisker2019,chechkin2009,chechkin2015}. Despite the generality and wide applicability of Eq.~\eqref{eq:genassrel}, it doesn't provide sufficient information regarding the  explicit form of $P_F(X,t)$ and how it decays with $X$.
 Our goal in this paper, using Large Deviations (LD) theory for random walks, is to show that a packet of generally non-Gaussian processes exhibits a universal decay for large $|X|$. 
 Our findings hold when thermal detailed balance holds (and so is the Crooks relation), but also beyond thermal settings.

The question regarding the properties of $P_F(X,t)$ for biased processes
is of great importance due to the
recent experimental and theoretical progress in addressing the universal exponential spread of positional PDF of single particles~\cite{Pinaki2007Universal,Wang2009Anomalous,Wang2012Brownian,Chubynsky2014Diffusing,Chechkin2017Brownian,ChechkinPRX2021}.
Besides a couple of exceptions~\cite{Chen2022,Weron2022}, all previous works focused on the unbiased case where $F=0$.
In particular, utilizing the continuous time random walk (CTRW)  when $F=0$, we explored theoretically exponential tails for the spreading packet of particles~\cite{BarkaiBurov2020,Wang2020Large}.
It was found~\cite{Touchette2009large,Pinaki2007Universal,Kege2000Direct,Masolivera200dynamic,Weeks2000Three,Leptos2009Dynamics,Eisenmann2010Shear,Toyota2011Non,Skaug2013Intermittent,Xue2016Probing,Jeanneret2016Entrainment,Chechkin2017Brownian,Cherstvy2019Non,Witzel2019Heterogeneities,Shin2019Anomalous,Singh2020Non,Mejia2020Tracer} that the tails of positional PDF of single particles in complex systems follow Laplace~\cite{Laplace1986} (exponential) and not Gaussian decay. Examples include phospholipid fluid tubules  and biofilament networks \cite{Wang2009Anomalous}, colloidal suspensions and glasses \cite{Weeks2000Three,Kege2000Direct,Pinaki2007Universal,Gao2009Intermittent,Greco2022}, molecular motion on a solid-liquid interface \cite{Skaug2013Intermittent,Wang2017Three}, polymer solutions \cite{Xue2016Probing}, living cells \cite{Munder2016transition}, and many other systems \cite{Wang2009Anomalous,Chechkin2017Brownian}. In all of these systems, the single particles/molecules are un-driven externally, i.e., $F=0$ as mentioned. 
What is to be
expected if the single molecules in these experiments
were driven? 
We first, by utilizing the CTRW framework \cite{Kotulski1995Asymptotic,Metzler2000random,Mainardi2004fractional,Burioni2014Scaling,Cairoli2015Anomalous,Kutner2017continuous,Morales2017Stochastic,Wang2020Fractional}, verify that the left-right asymmetry of Eq.~\eqref{eq:genassrel}, i.e., Crooks relation, is satisfied for an extensive class of processes where $P_F(X,t)$ is far from Gaussian and the transport is not diffusive, neither it is self-averaging \cite{bouchaud1992weak,rinn2000multiple,burov2007occupation,Akimoto2018Non}, or ergodic \cite{bouchaud1992weak,Yasmine2010Subdiffusion}. 
By extending the Crooks relation, we show how the asymmetric Laplace universality reveals itself in driven systems and how it is determined by the Laplace propagators present for undriven cases.

The widely applicable CTRW framework is as follows~\cite{Bouchaud1990Anomalous,Metzler2000random,Klafter2011FirstB,Berkowitz2006Modeling,Lefevere2011Large,Lefevre2021}: A particle starts from $x=0$ and after a period $\tau_1$ the first transition to some random position $x_1$ is generated according to the PDF $\phi_{0\to x_1}$.
After the second temporal period $\tau_2$ a second transition to some random position $x_2$ is drawn from $\phi_{x_1 \to x_2}$.  We assume that $\phi_{x_1 \to x_2}$ is fully defined by the position difference $x_2-x_1$, 
meaning that $\phi_{x_1\to x_2} = \phi_{0\to (x_2-x_1)}$.
The process repeats itself and is terminated when the total measurement time $t$ reaches a prescribed value. All the different $\tau_i$ are positively defined random variables distributed according to  $\psi(\tau)$. 
The standard random walk with discrete time-steps, i.e., $\forall i$ $\tau_i = c$, is achieved via $\psi(\tau_i) = \delta(\tau_i - c)$.
For the general case, the number of transitions  performed during $t$, $N_t$, is a random variable. The function $Q_t(N)$ is the probability to obtain $N_t$ transitions during time $t$   
and it depends on the specific form of $\psi(\tau)$~\cite{Bouchaud1990Anomalous,Metzler2000random,Burov2020Cond}. 
Due to independence of $\tau_i$ and $x_i$, we condition on the different outcomes of $N_t$ and write the positional PDF of $X=x_{N_t}$, $P_F(X,t)$, by using the subordination approach~\cite{Bouchaud1990Anomalous,Metzler2000random,Godreche2001Statistics,Burov2020Cond} 
\begin{equation}
P_F(X,t) = \sum_{N=0}^\infty {\cal P}_N(X) Q_t(N),
\label{eq:ctrwpxt}
\end{equation}
where ${\cal P}_N(X)$ is the positional PDF of $X$ given that exactly $N$ transitions were performed. 
The conditional PDF ${\cal P}_N(X)$ is provided by 
\begin{equation}
    \label{eq:pnxposition}
   {\cal P}_N(X) =
   \int_{-\infty}^{\infty}\dots\int_{-\infty}^{\infty} \phi_{0\to x_1}\dots \phi_{x_{N-1}\to X}\prod\limits_{i=1}^{N-1}dx_i,
   \end{equation}
while the integration is over all possible values of $x_1,x_2,\dots,x_{N-1}$.
The transition probabilities $\phi_{x\to x'}$ depend on the force $F$. 
If detailed balance conditions \cite{levin2017} are satisfied by $\phi_{x\to x'}$ then  $\phi_{x\to x'}\Big/\phi_{x'\to x} = \exp\left[F (x'-x)/k_BT\right]$. 
It is important to notice that the  detailed balance conditions imply that
\begin{equation}
    \label{eq:symmetryphi}
    \phi_{x\to x'} = f(x'-x) \exp\left(\frac{F(x'-x)}{2k_B T}\right),
\end{equation}
where $f(x)$ is a symmetric function, i.e., $f(x)=f(-x)$. 
If the detailed balance conditions hold, since $\phi_{0\to x}=\phi_{-x\to 0}$,  $\phi_{0\to x}=\phi_{0\to -x}\exp\left[Fx/k_BT\right]$ then obviously $\phi_{0\to x} \exp\left[-Fx/2k_BT\right]= \phi_{0\to -x} \exp\left[Fx/2k_BT\right]$. 
 Note that the results in this work also hold for systems where the temperature does not determine the dynamics (see below).

According to Eq.~\eqref{eq:symmetryphi} and the symmetry of $f(x)$,
\begin{equation}
    \label{eq:transition01}
    \begin{array}{l}
     \int_{-\infty}^{\infty}\dots\int_{-\infty}^{\infty} \phi_{0\to x_1}\dots \phi_{x_{N-1}\to X}\prod\limits_{i=1}^{N-1}dx_i=\\
    e^\frac{FX}{2k_BT}\int\dots\int \prod\limits_{i=1}^{N-1} f\left(-x_i-(-x_{i-1})\right)dx_i =
    e^{\frac{FX}{k_BT}}{\cal P}_N(-X)
      \end{array}
\end{equation}
and eventually Eq.~\eqref{eq:ctrwpxt} leads to
\begin{equation}
    \label{eq:finpxtcomp}
R_{FF}(X)=\ln\left[    \frac{P_F(X,t)}{P_F(-X,t)}\right] =
\frac{FX}{k_BT}
\end{equation}
i.e., the Crooks relation in Eq.~\eqref{eq:genassrel}.

\begin{figure}[t]
\centering
 \includegraphics[width=0.5\textwidth]{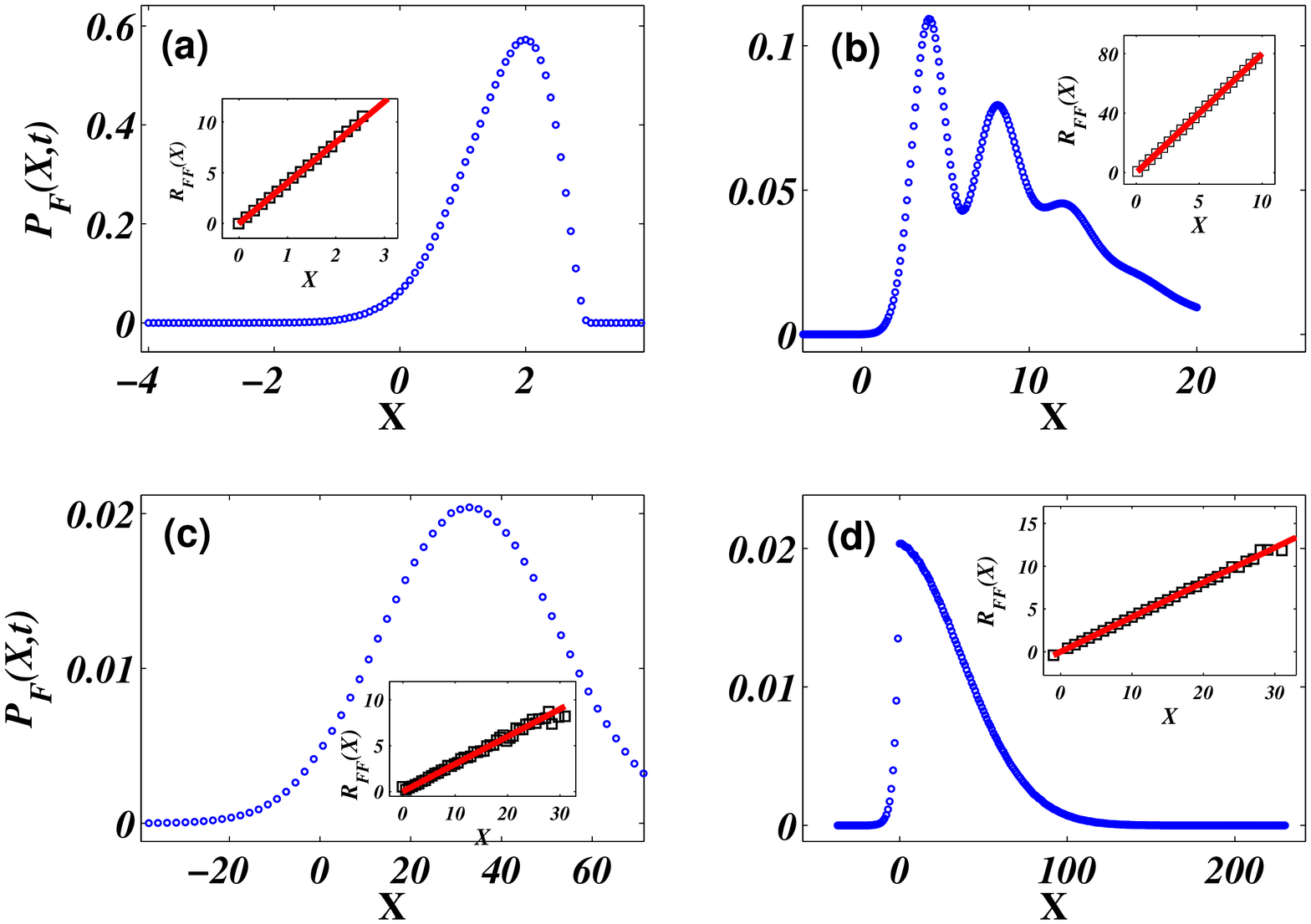}\\
 \caption{
 Positional PDF $P_F(X,t)$ for different models with $F\neq 0$. The insets display the validity of Crooks relation.
  Note that the shapes of the propagators seem non-universal. We will show that the far tails of a wide family of random walks belong to exponential, i.e., Laplace, universality.
 Symbols are numerical results while straight lines (in insets) follow  Eq.~\eqref{eq:genassrel}. The jump probabilities are of the form  $\phi_{x_1\to x_2}$ in Eq.~\eqref{eq:symmetryphi} and $\langle x \rangle = \int_{-\infty}^\infty x \phi_{0\to x}\,dx$ {\bf (a)}:  $f(x)=2/(e^{2}-e^{-2})$ ($x\in [-1,1]$), $\langle x \rangle=0.5373$, $\langle x \rangle F/k_BT=2.149$ and  $\psi(\tau)=\delta(\tau-1)$ ($t=3$). 
 {\bf(b)}  $f(x)=e^{-8} e^{-x^2/2}/\sqrt{2\pi}$, $\langle x \rangle = 4$, $\langle x \rangle F/k_BT=32$ and $\psi(\tau)=\exp(-\tau)$, ($t=2$).  {\bf (c)},  $f(x)=e^{-0.0225} e^{-x^2/2}/\sqrt{2\pi}$,  $\langle x \rangle = 0.1483$, $\langle x \rangle F/k_BT=0.0444$ and $\psi(\tau)=0.0474\times\tau^{-1-3/2}$
  ($t=10^3$). {\bf (d)} Quenched trap model, $f(x)=\delta(|x|-1)\sqrt{6/25}$, $\langle x \rangle = 1/5$, $\langle x \rangle F/k_BT=\ln(3/2)/5$,  the waiting times are quenched and distributed according to $\psi(\tau)=0.1581\tau^{-1-T/T_g}\Theta(\tau-0.1)$ with $T/T_g=1/2$ ($t=500$). See S.M. for details of the simulations.
}\label{fig:assymetryrelation}
\end{figure}

While Eq.~\eqref{eq:genassrel} was presented for Gaussian process, Eq.~\eqref{eq:finpxtcomp} holds for a much larger class of  non-Gaussian processes~\cite{seifert2007,esposito2008,bisker2019,chechkin2009,chechkin2015}.
In Fig.~\ref{fig:assymetryrelation}, the positional PDFs of several different processes are presented, and the insets show that Crooks relation is satisfied for all of them. 
In Fig.~\ref{fig:assymetryrelation} {\bf (a)} the PDF is measured at a finite $t$ before CLT conditions are reached (i.e., non-Gaussian propagator),  Eq.~\eqref{eq:finpxtcomp} is satisfied. The same happens for the oscillating form in Fig.~\ref{fig:assymetryrelation} {\bf (b)}. 
When $\psi(\tau)\sim \tau^{-1-\alpha}$ for $\tau\to\infty$ and $1<\alpha<2$, 
$P_F(X,t)$ is described by L\'{e}vy statistics (see~\cite{Bouchaud1990Anomalous,Wang2020Fractional}), nevertheless Eq.~\eqref{eq:finpxtcomp} is satisfied.
Even for cases when the disorder is quenched, and the behavior is non-self-averaging and anomalous~\cite{Burov2017,shafirBurov2022}, e.g., the Quenched trap model \cite{Bouchaud1990Anomalous,Burov2011Time,Akimoto2018Non},
Eq.~\eqref{eq:finpxtcomp}  is still applicable(see Fig.~\ref{fig:assymetryrelation} {\bf (d)}).

Figure~\ref{fig:assymetryrelation} shows that a diverse class of propagators satisfies the Crooks relation (Eq.~\eqref{eq:genassrel}). But this diversity also tells us that it is not feasible to use this relation to obtain explicit properties of $P_F(X,t)$. 
 Therefore our question regarding the decay properties of $P_F(X,t)$ still stands.     
To answer this, we now switch gears and address the connection between biased and unbiased propagators.
The asymmetry relation in Eq.~\eqref{eq:genassrel} suggests that the propagator $P_F(X,t)$ can be written as $P_F(X,t)=\exp(FX/2k_B T)H(X,t)$, where $H(X,t)$ is some symmetric function of $X$. 
What is $H(X,t)$, and can it be written in terms of a propagator for transport without external driving?

We start the search for representation of driven process in terms of undriven one by exploring a specific example: a driven process where $\psi(\tau)=\exp(-\tau/\langle \tau \rangle)/\langle \tau \rangle$ and $\phi_{x\to x'} =\exp\left(-[(x'-x)-b/2]^2/2\delta^2\right)/\sqrt{2\pi \delta^2} $, i.e., exponential distribution of the waiting times and biased Gaussian distribution of step sizes. 
If the system is in thermal equilibrium $b/\delta^2=F/k_B T$, however, our theory does not require this condition to hold, as some of the experimental systems mentioned in the introduction are active.
Taking the Fourier transform of Eq.~\eqref{eq:ctrwpxt} we obtain ${\tilde P}_F(k,t)=\sum_{N=0}^\infty \frac{1}{N!} (t/\langle \tau \rangle)^N e^{-t/\langle \tau \rangle} e^{-\frac{N}{2}(k\delta-i b/2\delta)^2-N b^2/8\delta^2}$. Then replacing $k\to-iu$ and summing the series results in 
${\tilde P}_F(u,t)= e^{K(u)}$ where $K(u)=-t/\langle \tau \rangle(1-e^{-b^2/8\delta^2}e^{\frac{1}{2}(u\delta+b/2\delta)^2})$. 
Next, by using the saddle point method, a well known Cramers-Daniels LD formula~\cite{Daniels1,Touchette2009large} is obtained $P_F(x,t)\sim e^{K(u^*)-u^*X}/\sqrt{2\pi K''(u^*)}$ while $u^*$ satisfies the equation $K'(u^*)=X$ that results in $u^*=-b/2\delta^2+\sqrt{W_0([\langle \tau \rangle X/t e^{-b^2/8\delta^2} \delta]^2)}/\delta$. The Lambert $W_0(z)$ function~\cite{Corless,Meerson,Lior2018} is the solution of the equation $ye^y = z$. Finally, we can write for the PDF the LD form in the large $|X|$ limit (see Supplemental Material (S.M.) for details)
\begin{equation}
    \label{eq:exmplPXT}
    P_F(X,t)\sim e^{-t I_F(X/t)}
\end{equation}
where
\begin{equation}
    \label{eq:exmplIone}
    I_F(z) = -\frac{b z}{2\delta^2}+\frac{1}{\langle \tau \rangle} + \frac{|z|}{\delta}\left[\sqrt{W_0([\frac{\langle \tau \rangle z}{ a \delta}]^2)}-\frac{1}{\sqrt{W_0([\frac{\langle \tau \rangle z}{ a \delta}]^2)}}\right]
\end{equation}
is the rate function and $a=e^{-b^2/8\delta^2}$. 
Due to the presence of the term $-b z/2\delta^2$ in $I_F(z)$, and since $W_0(z^2)\sim 2\ln(z)$ when $z>>1$, 
$P_F(X,t)$ attains asymmetric exponential decay (up to logarithmic corrections) for large $|X|$, i.e.,
\begin{equation}
\label{eq:pxtexmplass}
P_F(X,t) \underset{|X|\to\infty}{\sim} e^{\frac{b X}{2\delta^2}-\frac{1}{\sqrt{2}\delta}|X|\sqrt{\ln\left(\frac{\langle \tau \rangle}{\delta}\frac{|X|}{ at}\right)}}e^{-\frac{t}{\langle\tau\rangle}}.
\end{equation}
Eq.~\eqref{eq:pxtexmplass} proves that exponential tails, recorded in vast number of systems when $F=0$, can be found also for $F\neq 0$, however these tails are asymmetric. This indicates that 
 the universality of Laplace tails is far larger than known previously and holds for biased processes.

\begin{figure}[t]
 \centering
 \includegraphics[width=0.43\textwidth]{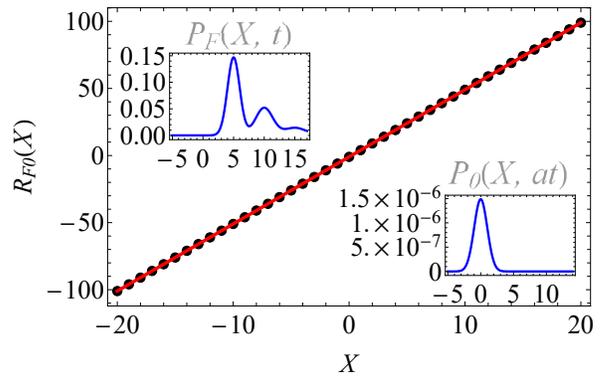}\\
 \caption{The connection between the PDF of a driven process and the PDF of an un-driven one, as described by Eq.~\eqref{eq:pxtexample}. The circles present $\ln \left[ P_F(X,t)/P_0(X,at)\right]$ while the thick line is the rhs of Eq.~\eqref{eq:pxtexample}. $\delta=\langle \tau \rangle =1$, $b=10$ and $t=1$. The insets describe $P_F(X,t)$ (presence of oscillations) and $P_0(X,at)$ (no oscillations are visible).
}\label{fig:linconect}
\end{figure}

 By taking $b= 0$ in Eq.~\eqref{eq:exmplIone} we obtain the rate function $I_0(z)$ that determines the PDF $P_0(X,t)\sim e^{-t I_0(X/t)}$ of an undriven process~\cite{Wang2020Large,Sokolov2021}. A simple connection between the rate functions is established 
 \begin{equation}
    \label{eq:ratefuncconect}
    I_F\left(z\right) = -\frac{b z}{2\delta^2} +\frac{1-a}{\langle \tau \rangle}+a I_0\left(\frac{z}{a}\right)
 \end{equation}
 that results in a connection between $P_F(X,t)$ and the propagator of undriven process at time $at$
\begin{equation}
    \label{eq:pxtexample}
    R_{F0}(X)=
  \ln\left[P_F(X,t)\Big/P_0(X,at)\right]
    = \frac{b X}{2\delta^2} -(1-a)\frac{t}{\langle\tau\rangle}.
\end{equation}
We show in S.M., that for the presented example of Gaussian $\phi_{x\to x'}$ and exponential $\psi(\tau)$, the equality in Eq.~\eqref{eq:pxtexample} holds for any $X$.

In the continuum limit, when the mean time between jumps becomes small and so is the step size (see S.M. for details), Eq.~\eqref{eq:pxtexample} converges to the $R_{F0}$ of a system in contact with a heat bath and for which Einstein relation is applicable. Meaning, $R_{F0}(X)=FX/2k_B T - DF^2 t/4(k_B T)^2$, when $P_F(X,t)=\exp\left[-(X-\frac{DFt}{k_B T})^2/4Dt\right]/\sqrt{4\pi Dt}$.     
 Unlike the Crooks relation, $R_{F0}(X)$ in Eq.~\eqref{eq:pxtexample} connects the PDF of a driven particle to the PDF of an undriven one. 
It is important to notice that temporal rescaling $t\to at$ of the unperturbed process is performed, i.e., the PDF $P_0$ is calculated at time $at$. 
In Fig.~\ref{fig:linconect} we present the case when $P_F(X,t)$ and $P_0(X,at)$ appear totally different while Eq.~\eqref{eq:pxtexample} holds $\forall X$.
The oscillations of $P_F(X,t)$ that are 
seen in 
the inset of Fig.~\ref{fig:linconect} are found in the regime when neither the Gaussian central limit theorem (large $t$) nor the LD theory (large $|X|$) in Eq.~\eqref{eq:pxtexmplass} hold.
In S.M. we show that there are extremely small modulations in the behavior of $P_0(x,t)$.
When $b$ is large, these negligible modulations become important due to the huge factor in the form of $e^{b X/2\delta^2}$, and therefore the application of $F$ serves as a "magnifying glass".

The established connection between the perturbed and unperturbed process, Eqs.~(\ref{eq:ratefuncconect}-\ref{eq:pxtexample}), and the asymmetric Laplace decay, can be extended for the case of more general  PDFs $\phi_{x\to x'}$ and $\psi(\tau)$. 
$\phi_{x\to x'}$ is normalized and hence  Eq.~\eqref{eq:symmetryphi} yields
$1=\int_{-\infty}^{\infty}f(x)\exp(Fx/2k_BT)\,dx=2\int_0^\infty f(x)\cosh(Fx/2k_BT)\,dx\geq \int_{-\infty}^{\infty}f(x)\,dx$. Accordingly, we can always write $\int_{-\infty}^{\infty}f(x)\,dx=\exp(-\lambda_F^2)$, where $\lambda_F$ is a strictly real constant that depends on $F$, explicitly $\lambda_F^2 = -\ln(\int_{-\infty}^{\infty} \phi_{0\to x}e^{-Fx/2k_BT}\,dx)$.
Notice that $\lambda_F \to 0$ when $F\to 0$ and for $F\to\infty$, $\lambda_F\to\infty$. For the Gaussian case discussed above, $\lambda_F^2=b^2/8\delta^2$.  
The probability of performing a jump of size $x$ for the unperturbed process, $f_0(x)$, is therefore   
\begin{equation}
    \label{eq:fxtofzero}
    f(x)=f_0(x)e^{-\lambda_F^2}.
\end{equation}
Then ${\cal P}_N(X)$ is written as 
    ${\cal P}_N(X)=e^{\frac{FX}{2 k_B T}} {\cal P}_{0,N}(X) e^{-\lambda_F^2 N}$,
where ${\cal P}_{0,N}(X)=\sum_{x_1,\dots,x_{N-1}}\prod_{i=1}^N f_0(x_i-x_{i-1})$, i.e., the PDF of the unperturbed process if exactly $N$ transitions were performed. 
Subsequently, the positional PDF is 
$P_F(X,t)=e^{FX/2k_B T}\sum_N {\cal P}_{0,N}(X) e^{-\lambda_F^2 N}Q_t(N)$. We focus on the large $|X|$ limit. Due to the detailed balance condition (Eq.~\eqref{eq:symmetryphi}),  $f(x)$ can't decay for large $|X|$ slower than exponentially. Therefore large displacements are achieved by performing a large number of transitions. Thus the limit of large $N$ for the behavior of $Q_t(N)$ must be addressed. 
Then for a finite $t$ only the short-$\tau$ properties of $\psi(\tau)$ are important~\cite{Pagnini2022} and we consider the case when $\psi(\tau)$ is analytic in the vicinity of $\tau=0$ \cite{BarkaiBurov2020}, i.e., 
\begin{equation}
    \label{eq:taylorpsitau}
    \psi(\tau)\sim
    C(1-\tau/\tau^*)
  \qquad   \tau\to 0,
\end{equation}
where $C>0$ and $\tau^*$ can be positive or negative (see discussion below). A more general Taylor expansion is addressed in S.M.
In Laplace space, due to convolution property of a sum, ${\hat Q}_s(N)=\int_0^\infty Q_t(N)e^{-t s}\,dt$, 
attains the form ${\hat Q}_s(N)={\hat \psi}(s)^N/s-{\hat \psi}(s)^{N+1}/s$, where ${\hat \psi}(s)$ is the Laplace transform of $\psi(\tau)$. In the limit when $t/N\to 0$, the inverse Laplace yields \cite{BarkaiBurov2020,Burov2020Cond}
\begin{equation}
    \label{eq:qtnasymptotic}
    Q_t(N)\sim\frac{\left(Ct\right)^{N}}{N!}\exp\left(-\frac{t}{\tau^*}\right).
\end{equation}
Eq.~\eqref{eq:qtnasymptotic} yields
\begin{equation}
    \label{eq:qtnrescaled}
    \exp(-N\lambda_F^2)Q_t(N)\underset{t/N\to 0}{\sim} \exp\left(-[1-{\tilde a}]\frac{t}{\tau^*}\right)Q_{{\tilde a}t}(N)
\end{equation}
where ${\tilde a}=\exp\left(-\lambda_F^2\right)=\int_{-\infty}^{\infty} \phi_{0\to x}e^{-Fx/2k_BT}\,dx$, i.e., the  moment generating function of $\phi_{0\to x}$, calculated at $-F/2k_BT$. 
Notice that $Q_{{\tilde a}t}(N)$ attains the form in Eq.~\eqref{eq:qtnasymptotic} when ${\tilde a}t/N\to 0$.   
Since the unperturbed PDF is $P_0(X,t)=\sum_N {\cal P}_N^0(X)Q_t(N)$, Eq.~\eqref{eq:qtnrescaled} yields 
\begin{equation}
    \label{eq:pxtgenform}
    R_{F0}(X)= \ln\left[P_F(X,t)\Big/P_0(X,{\tilde a}t)\right]\sim \frac{FX}{2k_B T}-[1-{\tilde a}]\frac{t}{\tau^*}
    \end{equation}
when $|X|\to\infty$.
In the language of LD and rate functions Eq.~\eqref{eq:pxtgenform} reads
\begin{equation}
I_F\left(z\right) = -\frac{Fz}{2K_BT}+\frac{1-{\tilde a}}{\tau^*}+{\tilde a}I_0\left(\frac{z}{{\tilde a}}\right)
    \label{eq:rategenform}
\end{equation}
when $z$ is large.

\begin{figure}[tp!]
 \centering
 \includegraphics[width=0.48\textwidth]{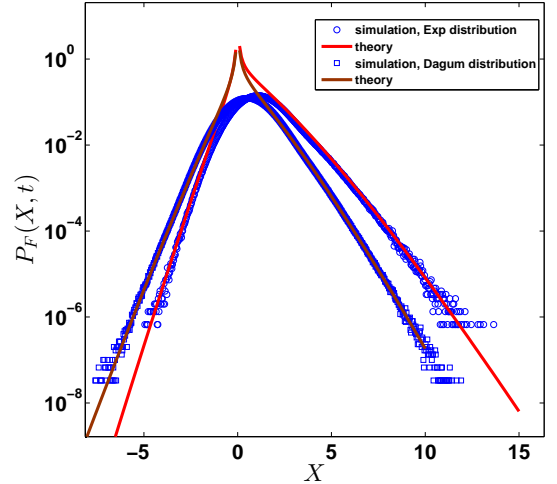}\vspace{-0.3cm}
 \caption{Universal asymmetric exponential tails of the positional distribution for different biases and various waiting time PDFs. We present
exponential distribution $\psi(\tau)=\exp(-\tau)$ (`$\square$') and Dagum distribution $\psi(\tau)=1/(1+\tau)^2$ (`$\circ$'). The corresponding displacement are drawn from Eq.~\eqref{eq:fxtofzero} and $f_0(x)=1/\sqrt{2\pi}\exp(-|x/\delta|^2)$ with $\lambda^2_F=1/2$, $F/K_BT=2$; and $\lambda^2_F=1/8$, $F/K_BT=1$, respectively.
 The symbols are simulations obtained from $10^8$ realizations with $t=0.5$ and the solid lines are theoretical predictions Eq.~\eqref{eq:pxtgenform} (see S.M. for details).
}\label{fig:pxttails}
\end{figure}

In the limit of small forces, $F\to 0$, ${\tilde a}\sim 1-\langle x \rangle F/2k_BT$, where $\langle x \rangle = \int_{-\infty}^{\infty}x\phi_{0\to x}\,dx$ is the average jump size. Then Eq.~\eqref{eq:pxtgenform} reads 
\begin{equation}
    \label{eq:linresponse}
    R_{F0}(X)\underset{F\to 0}{\sim}\frac{F}{2k_BT}\left(
    X-\langle x \rangle \frac{t}{\tau^*}
    \right).
\end{equation}
Equations~\eqref{eq:pxtgenform} and~\eqref{eq:linresponse} show  that the linear increase of $R_{F0}$ both in $X$ and $t$, found in Eq.~\eqref{eq:pxtexample}, is generally valid. 
In particular, the prefactor $F/2k_BT$ is universal.
 Eq.~\eqref{eq:linresponse} clearly shows that $|\tau^*|$, defined in Eq.~\eqref{eq:taylorpsitau}, is a novel and important microscopical time-scale that determines the dynamics of rare events, i.e. LD in $X$. While the usual assumption is that the average waiting time is the important microscopical time-scale, our results show that for large $X$, $|\tau^*|$ is the relevant time-scale to consider. 
 To reach large $X$, the particle must perform many jumps in a given time $t$. Therefore it typically spends only a short period of time at each location before performing a jump. Hence  only short-time properties of $\psi(\tau)$ are relevant. Notice that $\tau^*$ can be positive or negative. The sign of $\tau^*$ shows whether $\psi(\tau)$ is a decreasing/increasing function in the $\tau\to 0$ limit.

Asymmetric Laplace tails: Like for the case of Gaussian jumps and exponential waiting times, we focus on the  $|X|/(C \delta t)>>1$ limit of $P_0(X,t)$. When $f_0(|x|)\sim\exp(-|x/\delta|^\beta)$, for large $|x|$ (and $\beta>1$), ${\cal P}_N(X)$ is non-negligible (when $|X|$ is large) only for $N>>1$. In the large $|X|$ limit, the saddle point approximation can be applied to find the explicit form of $P_0(X,t)$ (see details in S.M. and~\cite{BarkaiBurov2020}).  
Then according to Eq.~\eqref{eq:pxtgenform} the large $|X|$ behavior of $P_F(X,t)$ is 
\begin{equation}
    \label{eq:pxtlargeX}
    P_F(X,t)\sim e^{\frac{FX}{2k_B T}-|X|\kappa\ln\left(\frac{|X|}{{C \delta \tilde a}t}\right)^{1-1/\beta}-\frac{t}{\tau^*}}.
\end{equation}
where $\kappa=\left(g_0/\beta+1/{g_0^{\beta-1}\delta^\beta}\right)\beta^{1-1/\beta}$ and $g_0=(\beta(\beta-1))^{1/\beta}/\delta$.

This result shows that the decay of the PDF of a driven particle/molecule is expected to be of exponential (i.e., Laplace) form for a broad range of processes.  In Fig.~\ref{fig:pxttails} this asymmetric exponential decay is displayed.
Only two conditions were exploited in the derivation of this result
(A) analytic form of $\psi(\tau)$ in the $\tau\to 0$ limit, Eq.~\eqref{eq:taylorpsitau} and (B) validity of Eq.~\eqref{eq:symmetryphi}.

 We have established two main results: (i) connection between $P_0(X,t)$ and $P_F(X,t)$ (and rate functions $I_0$, $I_F$);  (ii) asymmetric Laplace tails. (i) and (ii)  determine the response to an external field for an extensive class of processes. 
Our results are a strong statement regarding numerous experimental systems where single particles present non-Gaussian but rather Laplace behavior: Application of external force in such systems makes the PDF asymmetric (in accordance with Eq.~\eqref{eq:genassrel}) but does not change the nature of the Laplace tails (as prescribed by Eq.~\eqref{eq:pxtgenform}).
Moreover, the correspondence (Eq.~\eqref{eq:pxtgenform}) between a driven process and an un-driven process, measured at the shorter time (${\tilde a}$ always $<1$), provides the means of exploring the short-time behavior of an unperturbed process by applying external field.

{\bf{Acknowledgments:}}
This work was supported by the Israel Science Foundation Grants: 2796/20 (SB) and 1614/21 (EB).

\bibliographystyle{prestyle} 
\bibliography{./drivenparticlebib,wenxian} 

\begin{thebibliography}{10}

\bibitem{kubotoda}
R.~Kubo, M.~Toda, and N.~Hashitsume, \emph{Statistical Physics II:
  Nonequilibrium Statistical Mechanics} (Springer-Verlag, Heidelberg, 1998).

\bibitem{Bouchaud1990Anomalous}
J.-P. Bouchaud and A.~Georges, Anomalous diffusion in disordered media:
  Statistical mechanisms, models and physical applications, Phys. Rep.
  \textbf{195}, 127 (1990).

\bibitem{Crooks199}
G.~E. Crooks, Entropy production fluctuation theorem and the nonequilibrium
  work relation for free energy differences, Phys. Rev. E \textbf{60}, 2721
  (1999).

\bibitem{Crooks200}
G.~E. Crooks, Path-ensemble averages in systems driven far from equilibrium,
  Phys. Rev. E \textbf{61}, 2361 (2000).

\bibitem{bustamante2005}
D.~Collin, F.~Ritort, C.~Jarzynsky, S.~B. Smith, I.~T. Jr, and C.~Bustamante,
  Verification of the {C}rooks fluctuation theorem and recovery of rna folding
  free energies, Nature \textbf{437}, 231 (2005).

\bibitem{seifert2007}
T.~Speck and U.~Seifert, The {J}arzynski relation, fluctuation theorems, and
  stochastic thermodynamics for non-{M}arkovian processes, J. Stat. Mech.
  L09002 (2007).

\bibitem{esposito2008}
M.~Esposito and K.~Lindenberg, Continuous-time random walk for open systems:
  Fluctuation theorems and counting statistics, Phys. Rev. E \textbf{77},
  051119 (2008).

\bibitem{bisker2019}
I.~A. Martinez, G.~Bisker, J.~M. Horowitz, and J.~M.~R. Parrondo, Inferring
  broken detailed balance in the absence of observable currents, Nat. Comm.
  \textbf{10}, 3542 (2019).

\bibitem{chechkin2009}
A.~V. Chechkin and R.~Klages, Fluctuation relations for anomalous dynamics, J.
  Stat. Mech. L03002 (2009).

\bibitem{chechkin2015}
P.~Dieterich, R.~Klages, and A.~V. Chechkin, Fluctuation relations for
  anomalous dynamics generated by time-fractional {F}okker–{P}lanck
  equations, New J. Phys. \textbf{17}, 075004 (2015).

\bibitem{Pinaki2007Universal}
P.~Chaudhuri, L.~Berthier, and W.~Kob, Universal nature of particle
  displacements close to glass and jamming transitions, Phys. Rev. Lett.
  \textbf{99}, 060604 (2007).

\bibitem{Wang2009Anomalous}
B.~Wang, S.~M. Anthony, S.~C. Bae, and S.~Granick, Anomalous yet {B}rownian,
  Proc. Natl. Acad. Sci. U.S.A. \textbf{106}, 15160 (2009).

\bibitem{Wang2012Brownian}
B.~Wang, J.~Kuo, S.~C. Bae, and S.~Granick, When brownian diffusion is not
  gaussian, Nat. Mater. \textbf{11}, 481 (2012).

\bibitem{Chubynsky2014Diffusing}
M.~V. Chubynsky and G.~W. Slater, Diffusing diffusivity: A model for anomalous,
  yet {B}rownian, diffusion, Phys. Rev. Lett. \textbf{113}, 098302 (2014).

\bibitem{Chechkin2017Brownian}
A.~V. Chechkin, F.~Seno, R.~Metzler, and I.~M. Sokolov, Brownian yet
  non-{G}aussian diffusion: From superstatistics to subordination of diffusing
  diffusivities, Phys. Rev. X \textbf{7}, 021002 (2017).

\bibitem{ChechkinPRX2021}
J.~M. Miotto, S.~Pigolotti, A.~V. Chechkin, and S.~Rold\'an-Vargas, Length
  scales in {B}rownian yet non-{G}aussian dynamics, Phys. Rev. X \textbf{11},
  031002 (2021).

\bibitem{Chen2022}
X.~Wang and Y.~Chen, Random diffusivity processes in an external force field,
  Phys. Rev. E \textbf{106}, 024112 (2022).

\bibitem{Weron2022}
Y.~Lanoisele'e, A.~S. Stanislavsky, D.~Calebiro, and A.~Weron, Temperature and
  friction fluctuations inside a harmonic potential, arXiv:2207.14068  (2022).

\bibitem{BarkaiBurov2020}
E.~Barkai and S.~Burov, Packets of diffusing particles exhibit universal
  exponential tails, Phys. Rev. Lett. \textbf{124}, 060603 (2020).

\bibitem{Wang2020Large}
W.~Wang, E.~Barkai, and S.~Burov, Large deviations for continuous time random
  walks, Entropy \textbf{22}, 697 (2020).

\bibitem{Touchette2009large}
H.~Touchette, The large deviation approach to statistical mechanics, Phys. Rep.
  \textbf{478}, 1 (2009).

\bibitem{Kege2000Direct}
W.~K. Kegel and A.~van Blaaderen, Direct observation of dynamical
  heterogeneities in colloidal hard-sphere suspensions, Science \textbf{287},
  290 (2000).

\bibitem{Masolivera200dynamic}
J.~Masoliver, M.~Montero, and J.~M. Porr{\`a}, A dynamical model describing
  stock market price distributions, Physica A \textbf{283}, 559 (2000).

\bibitem{Weeks2000Three}
E.~R. Weeks, J.~C. Crocker, A.~C. Levitt, A.~Schofield, and D.~A. Weitz,
  Three-dimensional direct imaging of structural relaxation near the colloidal
  glass transition, Science \textbf{287}, 627 (2000).

\bibitem{Leptos2009Dynamics}
K.~C. Leptos, J.~S. Guasto, J.~P. Gollub, A.~I. Pesci, and R.~E. Goldstein,
  Dynamics of enhanced tracer diffusion in suspensions of swimming eukaryotic
  microorganisms, Phys. Rev. Lett. \textbf{103}, 198103 (2009).

\bibitem{Eisenmann2010Shear}
C.~Eisenmann, C.~Kim, J.~Mattsson, and D.~A. Weitz, Shear melting of a
  colloidal glass, Phys. Rev. Lett. \textbf{104}, 035502 (2010).

\bibitem{Toyota2011Non}
T.~Toyota, D.~A. Head, C.~F. Schmidt, and D.~Mizuno, Non-{G}aussian athermal
  fluctuations in active gels, Soft Matter \textbf{7}, 3234 (2011).

\bibitem{Skaug2013Intermittent}
M.~J. Skaug, J.~Mabry, and D.~K. Schwartz, Intermittent molecular hopping at
  the solid-liquid interface, Phys. Rev. Lett. \textbf{110}, 256101 (2013).

\bibitem{Xue2016Probing}
C.~Xue, X.~Zheng, K.~Chen, Y.~Tian, and G.~Hu, Probing non-{G}aussianity in
  confined diffusion of nanoparticles, J. Phys. Chem. \textbf{7}, 514 (2016).

\bibitem{Jeanneret2016Entrainment}
R.~Jeanneret, D.~O. Pushkin, V.~Kantsler, and M.~Polin, Entrainment dominates
  the interaction of microalgae with micron-sized objects, Nat. Commun.
  \textbf{7}, 12518 (2016).

\bibitem{Cherstvy2019Non}
A.~G. Cherstvy, S.~Thapa, C.~E. Wagner, and R.~Metzler, Non-{G}aussian{,}
  non-ergodic{,} and non-{F}ickian diffusion of tracers in mucin hydrogels,
  Soft Matter \textbf{15}, 2526 (2019).

\bibitem{Witzel2019Heterogeneities}
P.~Witzel, M.~G{\"o}tz, Y.~Lanoisel{\'e}e, T.~Franosch, D.~S. Grebenkov, and
  D.~Heinrich, Heterogeneities shape passive intracellular transport, Biophys.
  J. \textbf{117}, 203 (2019).

\bibitem{Shin2019Anomalous}
K.~Shin, S.~Song, Y.~H. Song, S.~Hahn, J.-H. Kim, G.~Lee, I.-C. Jeong, J.~Sung,
  and K.~T. Lee, Anomalous dynamics of in vivo cargo delivery by motor protein
  multiplexes, J. Phys. Chem. \textbf{10}, 3071 (2019).

\bibitem{Singh2020Non}
R.~K. Singh, J.~Mahato, A.~Chowdhury, A.~Sain, and A.~Nandi, Non-{G}aussian
  subdiffusion of single-molecule tracers in a hydrated polymer network, J.
  Chem. Phys. \textbf{152}, 024903 (2020).

\bibitem{Mejia2020Tracer}
C.~Mejia-Monasterio, S.~Nechaev, G.~Oshanin, and O.~Vasilyev, Tracer diffusion
  on a crowded random manhattan lattice, New J. Phys. \textbf{22}, 033024
  (2020).

\bibitem{Laplace1986}
P.~S. Laplace, Memoir on the probability of the causes of events, Statistical
  Science \textbf{1}, 364 (1986).

\bibitem{Gao2009Intermittent}
Y.~Gao and M.~L. Kilfoil, Intermittent and spatially heterogeneous
  single-particle dynamics close to colloidal gelation, Phys. Rev. E
  \textbf{79}, 051406 (2009).

\bibitem{Greco2022}
F.~Rusciano, R.~Pastore, and F.~Greco, Fickian non-gaussian diffusion in
  glass-forming liquids, Phys. Rev. Lett. \textbf{128}, 168001 (2022).

\bibitem{Wang2017Three}
D.~Wang, H.~Wu, and D.~K. Schwartz, Three-dimensional tracking of interfacial
  hopping diffusion, Phys. Rev. Lett. \textbf{119}, 268001 (2017).

\bibitem{Munder2016transition}
M.~C. Munder, D.~Midtvedt, T.~Franzmann, E.~N\"{u}ske, O.~Otto, M.~Herbig,
  E.~Ulbricht, P.~M\"{u}ller, A.~Taubenberger, S.~Maharana, L.~Malinovska,
  D.~Richter, J.~Guck, V.~Zaburdaev, and S.~Alberti, A p{H}-driven transition
  of the cytoplasm from a fluid- to a solid-like state promotes entry into
  dormancy, eLife \textbf{5}, e09347 (2016).

\bibitem{Kotulski1995Asymptotic}
M.~Kotulski, Asymptotic distributions of continuous-time random walks: A
  probabilistic approach, J. Stat. Phys. \textbf{81}, 777 (1995).

\bibitem{Metzler2000random}
R.~Metzler and J.~Klafter, The random walk's guide to anomalous diffusion: a
  fractional dynamics approach, Phys. Rep. \textbf{339}, 1 (2000).

\bibitem{Mainardi2004fractional}
F.~Mainardi, R.~Gorenflo, and E.~Scalas, A fractional generalization of the
  {P}oisson processes, Vietnam J. Math. \textbf{32}, 53 (2004).

\bibitem{Burioni2014Scaling}
R.~Burioni, G.~Gradenigo, A.~Sarracino, A.~Vezzani, and A.~Vulpiani, Scaling
  properties of field-induced superdiffusion in continuous time random walks,
  Commun. Theor. Phys. \textbf{62}, 514 (2014).

\bibitem{Cairoli2015Anomalous}
A.~Cairoli and A.~Baule, Anomalous processes with general waiting times:
  Functionals and multipoint structure, Phys. Rev. Lett. \textbf{115}, 110601
  (2015).

\bibitem{Kutner2017continuous}
R.~Kutner and J.~Masoliver, The continuous time random walk, still trendy:
  fifty-year history, state of art and outlook, Eur. Phys. J. B \textbf{90}, 50
  (2017).

\bibitem{Morales2017Stochastic}
V.~L. Morales, M.~Dentz, M.~Willmann, and M.~Holzner, Stochastic dynamics of
  intermittent pore-scale particle motion in three-dimensional porous media:
  Experiments and theory, Geophys. Res. Lett. \textbf{44}, 9361 (2017).

\bibitem{Wang2020Fractional}
W.~Wang and E.~Barkai, Fractional advection-diffusion-asymmetry equation, Phys.
  Rev. Lett. \textbf{125}, 240606 (2020).

\bibitem{bouchaud1992weak}
J.-P. Bouchaud, Weak ergodicity breaking and aging in disordered systems,
  Journal de Physique I \textbf{2}, 1705 (1992).

\bibitem{rinn2000multiple}
B.~Rinn, P.~Maass, and J.-P. Bouchaud, Multiple scaling regimes in simple aging
  models, Phys. Rev. Lett. \textbf{84}, 5403 (2000).

\bibitem{burov2007occupation}
S.~Burov and E.~Barkai, Occupation time statistics in the quenched trap model,
  Phys. Rev. Lett. \textbf{98}, 250601 (2007).

\bibitem{Akimoto2018Non}
T.~Akimoto, E.~Barkai, and K.~Saito, Non-self-averaging behaviors and
  ergodicity in quenched trap models with finite system sizes, Phys. Rev. E
  \textbf{97}, 052143 (2018).

\bibitem{Yasmine2010Subdiffusion}
Y.~Meroz, I.~M. Sokolov, and J.~Klafter, Subdiffusion of mixed origins: When
  ergodicity and nonergodicity coexist, Phys. Rev. E \textbf{81}, 010101(R)
  (2010).

\bibitem{Klafter2011FirstB}
J.~Klafter and I.~M. Sokolov, \emph{First Steps in Random Walks: From Tools to
  Applications} (Oxford University Press, Oxford, 2011).

\bibitem{Berkowitz2006Modeling}
B.~Berkowitz, A.~Cortis, M.~Dentz, and H.~Scher, Modeling non-fickian transport
  in geological formations as a continuous time random walk, Rev. Geophys.
  \textbf{44} (2006).

\bibitem{Lefevere2011Large}
R.~Lefevere, M.~Mariani, and L.~Zambotti, Large deviations for renewal
  processes, Stoch. Process. Their Appl. \textbf{121}, 2243 (2011).

\bibitem{Lefevre2021}
H.~Horii, R.~Lefevere, and T.~Nemoto, Large time asymptotic of heavy tailed
  renewal processes, J. Stat. Phys. \textbf{186}, 11 (2022).

\bibitem{Burov2020Cond}
S.~Burov, Limit forms of the distribution of the number of renewals,
  arXiv:2007.00381v1  (2020).

\bibitem{Godreche2001Statistics}
C.~Godr{\`e}che and J.~M. Luck, Statistics of the occupation time of renewal
  processes, J. Stat. Phys. \textbf{104}, 489 (2001).

\bibitem{levin2017}
D.~A. Levin and Y.~Peres, \emph{Markov Chains and Mixing Times} (American
  Mathematical Society, Providence, 2017).

\bibitem{Burov2017}
S.~{Burov}, {From quenched disorder to continuous time random walk}, \pre
  \textbf{96}, 050103(R) (2017).

\bibitem{shafirBurov2022}
D.~Shafir and S.~Burov, The case of the biased quenched trap model in two
  dimensions with diverging mean dwell times, J. Stat. Mech. \textbf{3}, 033301
  (2022).

\bibitem{Burov2011Time}
S.~Burov and E.~Barkai, Time transformation for random walks in the quenched
  trap model, Phys. Rev. Lett. \textbf{106}, 140602 (2011).

\bibitem{Daniels1}
H.~E. Daniels, Saddlepoint approximations in statistics, Ann. Math. Stat.
  \textbf{25}, 631 (1954).

\bibitem{Corless}
R.~M. Corless, G.~H. Gonnet, D.~E.~G. Hare, D.~J. Jeffrey, and D.~E. Knuth, On
  the {L}ambert{W} function, Advances in Computational Mathematics \textbf{5},
  329 (1996).

\bibitem{Meerson}
L.~Zarfaty and B.~Meerson, Statistics of large currents in the
  {K}ipnis-{M}archioro-{P}resutti model in a ring geometry, J. Stat. Mech.
  033304 (2016).

\bibitem{Lior2018}
L.~Zarfaty, A.~Peletskyi, I.~Fouxon, S.~Denisov, and E.~Barkai, Dispersion of
  particles in an infinite-horizon {L}orentz gas, Phys. Rev. E \textbf{98},
  010101(R) (2018).

\bibitem{Sokolov2021}
A.~Pacheco-Pozo and I.~M. Sokolov, Large deviations in continuous-time random
  walks, Phys. Rev. E \textbf{103}, 042116 (2021).

\bibitem{Pagnini2022}
S.~Vitali, P.~Paradisi, and G.~Pagnini, Anomalous diffusion originated by two
  {M}arkovian hopping-trap mechanisms, J. Phys. A:Math. Theor. \textbf{55},
  224012 (2022).

\end{thebibliography}

\pagebreak
\widetext
\begin{center}
\textbf{\large Supplemental Material for ``Laplace Tails and Asymmetry Relations for the Spread of Biased Random Walks''}
\end{center}
\setcounter{equation}{0}
\setcounter{figure}{0}
\setcounter{table}{0}
\setcounter{page}{1}
\makeatletter
\renewcommand{\theequation}{S\arabic{equation}}
\renewcommand{\thefigure}{S\arabic{figure}}


\section{Simulations Details for F\MakeTextLowercase{ig}.~1 of the main text}
In Fig.~1 in the main text, we consider several processes, obeying Eq.~(6). All along the manuscript, the initial position of the particles is a delta function, i.e., $\delta(x)$, and all the displacement of the particles are independent and identically distributed (IID) random variables. The details of the subplots of Fig.~1 are as follows:
\begin{itemize}
  \item For Fig.~1~(a), we consider standard random walk, where the waiting times of the random walkers are fixed, i.e., the waiting times follows $\psi(\tau)=\delta(\tau-1)$. In our simulation the observation time $t=3$ and this indicates that each trajectory of the particles has three renewals. Note that the displacement $x_i$ follows Eq.~(2) with $F/k_BT=4$ and $f(x)=2/(\exp(2)-\exp(-2))$ for $x\in[-1,1]$. Thus, the final position of the particles is $x_1+x_2+x_3$.
  \item In Fig.~1 (b) and (c), our model is the continuous time random walk (CTRW) model \cite{Bouchaud1990Anomalous,Metzler2000random,Klafter2011FirstB} as described in the main text.
  For (b) the waiting times distribution $\psi(\tau)$ is exponential and the distribution of jumps is Gaussian. Measurement time in (b) is $t=3$ (a.u). For (c) the waiting times distribution $\psi(\tau)$ is power-law and the distribution of jumps is Gaussian. The measurement time is $t=10^3$ (a.u).  
  \item For Fig.~1~(d), the Quenched trap model~\cite{Bouchaud1990Anomalous} is investigated. Here the particle is assumed to walk in the one-dimensional lattice. The lattice points are $\{-La,-(L-1)a,\cdots, -2a, -a, 0, a, 2a, \cdots, (L-1)a, La\}$ with integer $L\to +\infty$ and lattice
spacing $a=1$. On each lattice point a random energy $E_x$
is assigned, which is minus the energy of the particle on
site $x$, so $E_x>0$ is the depth of a trap on site $x$.  When the distribution of $E_x$ is $g(E_x)=\exp(-E_x/T_g)/T_g$, the average escape time is distributed according to
\begin{equation}\label{phiTaux}
\phi(\tau_x)\sim \alpha \tau_0^\alpha \tau_x^{-\alpha-1}, \tau_x\geq \tau_0
\end{equation}
with $\tau_x\to+\infty$ and $\alpha= T/T_g$, where $T<T_g$. Once the waiting time for specific lattice point was generated according to Eq.~\eqref{phiTaux}, the particle has to be trapped for the same waiting time $\tau_x$ every time  it visits this position. In that sense, compared with continuous time random walk, here the waiting times of each lattice point $\tau_x$ are position-dependent. 
For the displacement, we assume that the particle moves to right direction with probability $0.6$ and the left direction with $0.4$, respectively. In (d) we use $T/T_g=0.5$ and the measurement time is $t=500$ (a.u).
\end{itemize}


\section{Example of Asymmetric Exponential tails}\label{sect121}
\subsection{ $P_F(X,t)$ and $P_{0}(X,t)$ for the case of exponential $\psi(\tau)$ and Gaussian $\phi_{x\to x'}$}

As mentioned in the main text, in real space, the formal solution of the positional distribution of the CTRW model \cite{Metzler2000random} is
\begin{equation}\label{SMeq101}
P_F(X,t)=\sum_{N=0}^{\infty}Q_t(N)\mathcal{P}_N(X),
\end{equation}
where $Q_t(N)$ is the PDF of number of renewals and $\mathcal{P}_N(X)$ describes the PDF of $X$ provided exactly $N$ steps were made up to the observation time $t$. From the renewal theory, in Laplace space, the PDF of $N$ takes the form \cite{Godreche2001Statistics}
\begin{equation}\label{SMeq102}
\hat{Q}_s(N)=\hat{\psi}^N(s)\frac{1-\hat{\psi}(s)}{s}.
\end{equation}
Here $\hat{Q}_s(N)$ and $\hat{\psi}(s)$ denote the Laplace transforms of $Q_t(N)$ and $\psi(t)$, respectively.

In this example $\psi(\tau)=\exp(-\tau/\langle\tau\rangle)/\langle\tau\rangle$, i.e., exponential distribution of the waiting times.  
From Eq.~\eqref{SMeq102} the PDF of the number of renewals follows a Poisson distribution
\begin{equation}\label{SMeq201}
Q_t(N)=\frac{1}{N!}\exp\left(-\frac{t}{\langle\tau\rangle}\right)\left(\frac{t}{\langle\tau\rangle}\right)^N.
\end{equation}
For the spatial transitions we consider Gaussian distribution
\begin{equation}\label{GausssianDis}
\phi_{0\to x}=\frac{1}{\sqrt{2\pi\delta^2}}\exp\left(-\frac{(x-b/2)^2}{2\delta^2}\right).
\end{equation}
The conditional PDF $\mathcal{P}_N(X)$ to occupy location $X$ after $N$ transitions  follows
\begin{equation}\label{SMeq202}
\mathcal{P}_N(X)=\frac{1}{\sqrt{2\pi N\delta^2}}\exp\left(-\frac{(X-bN/2)^2}{2N\delta^2}\right).
\end{equation}
According to Eq.~\eqref{SMeq101}, $P_F(X,t)$ is
\begin{equation}\label{SMeq20301}
\begin{split}
P_F(X,t)&=\sum_{N=0}^\infty \exp\left(-\frac{t}{\langle\tau\rangle}\right)\frac{(t/\langle\tau\rangle)^N}{N!}\frac{1}{\sqrt{2\pi N\delta^2}}\exp\left(-\frac{(X-bN/2)^2}{2N\delta^2}\right).
\end{split}
\end{equation}
By rearranging the terms in Eq.~\eqref{SMeq20301} we determine the PDF $P_F(X,t)$ in terms of the positional PDF when the $b=0$ (no force) $P_0(X,t)$,
\begin{equation}\label{SMeq2030111}
\begin{split}
P_F(X,t)&=\exp\left(\frac{bX}{2\delta^2}\right)\exp\left(-\left(1-\exp\left(-\frac{b^2}{8\delta^2}\right)\right)\frac{t}{\langle\tau\rangle}\right)P_0\left(X,\exp\left(-\frac{b^2}{8\delta^2}\right)t\right),
\end{split}
\end{equation}
where
\begin{equation}\label{SMeqnonbiase}
\begin{split}
P_0(X,t)&=\sum_{N=0}^{\infty} \exp\left(-\frac{t}{\langle\tau\rangle}\right)\frac{(t/\langle\tau\rangle)^N}{N!}\frac{1}{\sqrt{2\pi N\delta^2}}\exp\left(-\frac{X^2}{2N\delta^2}\right).
\end{split}
\end{equation}
From Eq.~\eqref{SMeqnonbiase} we obtain Eq. (11) of the main text, that holds for any $X$ and $t>0$
\begin{equation}\label{ewssj101}
R_{F0}(X) = \ln\left[\frac{P_F(X,t)}{P_0(X,at)}\right]=\frac{bX}{2\delta^2}-\left(1-\exp\left(-\frac{b^2}{8\delta^2}\right)\right)\frac{t}{\langle\tau\rangle},
\end{equation}
where $a=\exp\left(-\frac{b^2}{8\delta^2}\right)$.
The continuum limit is defined by taking to $0$ the sizes of the steps $\delta\to 0$, and also the average waiting time at each position $\langle \tau \rangle\to 0$. At the same time $D=\delta^2/2\langle \tau \rangle$, i.e., the diffusion constant is kept constant. The force is introduced via $b/\delta^2=F/k_BT$. Therefore in the limit of weak forces $\exp(-b^2/8\delta^2)\to 1$ and $(1-\exp(-b^2/8\delta^2))/\langle \tau \rangle \to DF^2/4(k_BT)^2$.   
Thus in this limit 
\begin{equation}
    \label{rf0BM}
    R_{F0}(X) = FX/2k_B T - DF^2 t/4(k_B T)^2.
\end{equation}
For the case of a system coupled to a heat bath that describes Brownian motion, the propagator $P_0(X,t) = \exp\left[-X^2/4Dt\right]/\sqrt{4\pi Dt}$. According to Einstein relation in the limit of weak forces  $P_F(X,t) = \exp\left[-(X-DFT/k_Bt)^2/4Dt\right]/\sqrt{4\pi Dt}$, and therefore the continuum limit of $R_{F0}$ (Eq.~\eqref{rf0BM}) coincides with $R_{F0}$ for Brownian motion.

\begin{figure}[tbp]
  \centering
  \includegraphics[width=0.5\linewidth]{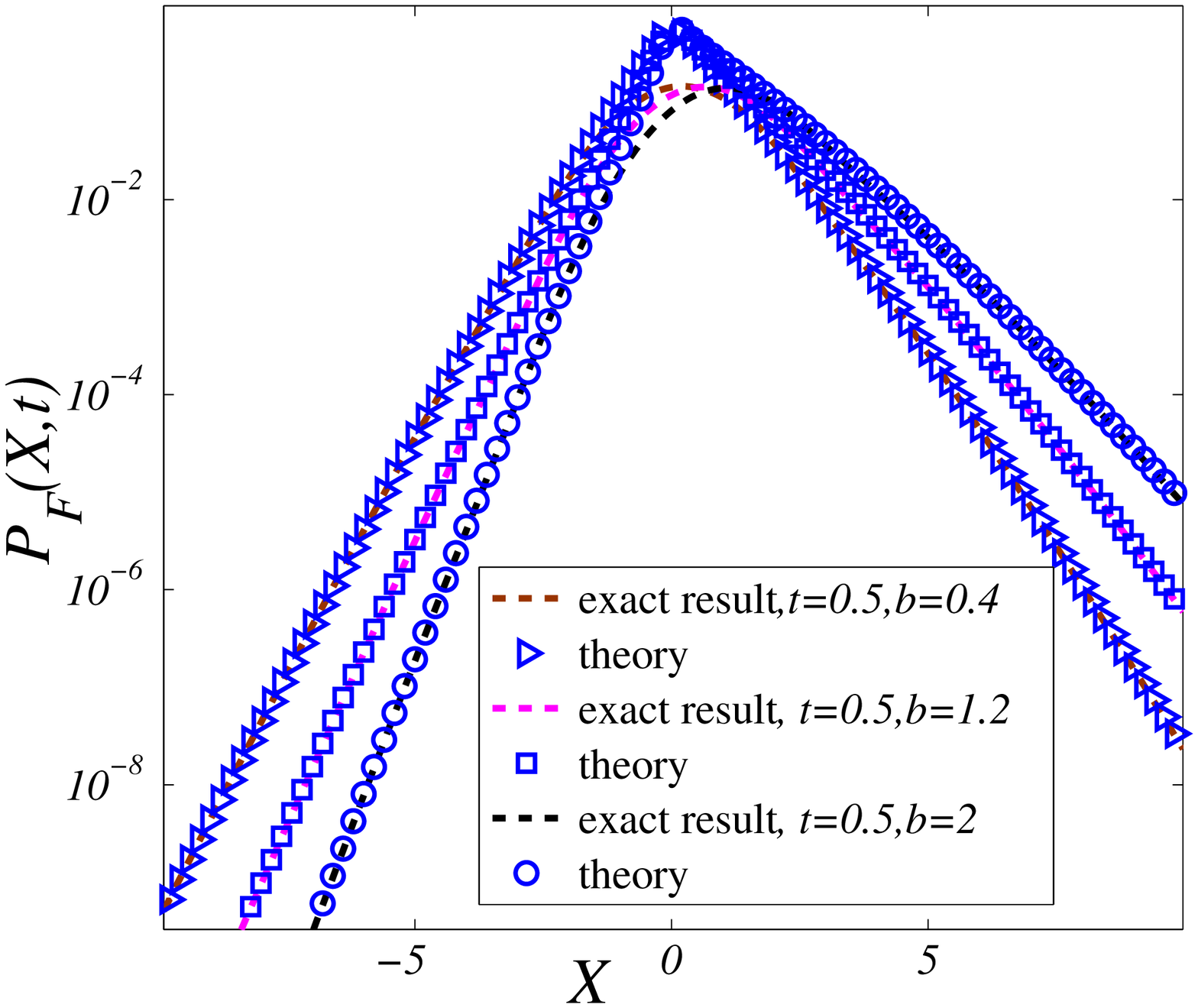}\\
  \caption{Exponential decay of the far tails of the positional PDF for a driven system. Here the waiting time follows $\psi(\tau)=\exp(-\tau)$ and the displacements are drawn from Gaussian distribution (see Eq.~\eqref{GausssianDis}) with $\delta=1$. We choose $t=0.5$. The exact result are obtained from the first line of Eq.~\eqref{SMeq20301} and the theoretical prediction Eq.~\eqref{SMeq2030111} with $P_0(X,t)$ calculated from Eq.~\eqref{SMeq20512}.    }
 \label{ExpGaussianPXT}
\end{figure}

\subsubsection{Decay of $P_F(X,t)$}

Below, we explore the tails of the  asymmetric positional distribution using symmetric one, i.e., Eq.~\eqref{SMeq2030111}. Rewriting $P_0(X,t)$, and using an integral approximation for the sum, we obtain
\begin{equation}\label{SMeq203}
\begin{split}
P_0(X,t)&=\sum_{N=0}^\infty Q_t(N) \mathcal{P}_{0,N}(X)\\
&\rightarrow \int_0^\infty \exp\left(-\frac{t}{\langle\tau\rangle}\right)\frac{(t/\langle\tau\rangle)^N}{N!}\frac{1}{\sqrt{2\pi N\delta^2}}\exp\left(-\frac{X^2}{2N\delta^2}\right) d N,
\end{split}
\end{equation}
where $N$ is treated as a continuous variable. 
The integral in Eq.~\eqref{SMeq203}, for the case when $|X|$ is large, is determined by large $N$ behavior of the integrand.  
Using Stirling's approximation $N!\sim \sqrt{2\pi N}(N/e)^N$ and rewriting Eq.~\eqref{SMeq203}, we obtain
\begin{equation}\label{SMeq204}
P_0(X,t)\sim \int_0^\infty \exp\left(-\frac{X^2}{2 \delta^2 N}-\ln (2 \pi  \delta N)+N \ln \left(\frac{t}{\langle\tau\rangle}\right)-\frac{t}{\langle\tau\rangle}-N \ln \left(\frac{N}{e}\right)\right)d N.
\end{equation}
Based on the saddle point approximation \cite{Daniels1}, Eq.~\eqref{SMeq204} yields
\begin{equation}\label{SMeq20512}
P_0(X,t)\underset{|X|\to\infty}{\sim} \frac{\exp \left(-|X|\left(\frac{\ln \left(\frac{\langle\tau\rangle | X| }{e \delta  t \sqrt{W_0\left(\frac{\langle\tau\rangle^2 X^2}{\delta ^2 t^2}\right)}}\right)}{\delta  \sqrt{W_0\left(\frac{\langle\tau\rangle^2 X^2}{\delta^2 t^2}\right)}}+\frac{\sqrt{W_0\left(\frac{\langle\tau\rangle^2 X^2}{\delta ^2 t^2}\right)}}{2 \delta }\right)-\frac{t}{\langle\tau\rangle}\right)}{\sqrt{2 \pi } \sqrt{\left| \delta  X \left(\sqrt{W_0\left(\frac{\langle\tau\rangle^2 X^2}{\delta ^2 t^2}\right)}+\frac{1}{\sqrt{W_0\left(\frac{\langle\tau\rangle^2 X^2}{\delta ^2 t^2}\right)}}\right)\right| }}.
\end{equation}
See also recent works in Refs. \cite{BarkaiBurov2020,Wang2020Large}.
Here $W_0$ is called the Lambert $W_0$ function \cite{Corless}, and the subscript zero describes the branch of this function. Specifically, $y=W_0(x)$ satisfies the equation $x=ye^y$. The theoretical results are verified by comparison to exact results in Fig.~\ref{ExpGaussianPXT}. 
If $ (|X|/\delta)/(t/\langle \tau\rangle)\to \infty$,
the asymptotic behavior of Eq.~\eqref{SMeq20512} yields
\begin{equation}\label{P000XT}
P_0(X,t)\sim \exp \left(-\frac{|X|}{2\delta}\left( \sqrt{\ln \left(\frac{\langle\tau\rangle^2 X^2}{\delta ^2 t^2}\right)}\right)-\frac{t}{\langle\tau\rangle}\right),
\end{equation}
where we used the relation $W_0(|x|)\propto \ln(|x|)$ for large $|x|$.
Eq.~\eqref{P000XT} show that the decay of the positional PDF in the un-biased case $(b=0)$ is exponential up to logarithmic corrections. 
According to Eqs.~\eqref{SMeq2030111} and \eqref{SMeq20512}, the far tails of the positional distribution for a driven system are provided by
\begin{equation}\label{eq:unbiasedLong}
P_F(X,t)\sim \exp \left(\frac{bX}{2\delta^2}-\frac{t}{\langle\tau\rangle}-\frac{|X|}{2\delta}\left( \sqrt{\ln \left(\frac{\langle\tau\rangle^2 X^2}{\delta ^2 (\exp\left(-\frac{b^2}{8\delta^2}\right)t)^2}\right)}\right)\right),
\end{equation}

\subsubsection{Strong Bias}

\begin{figure}[tbp]
  \centering
  \includegraphics[width=0.5\linewidth]{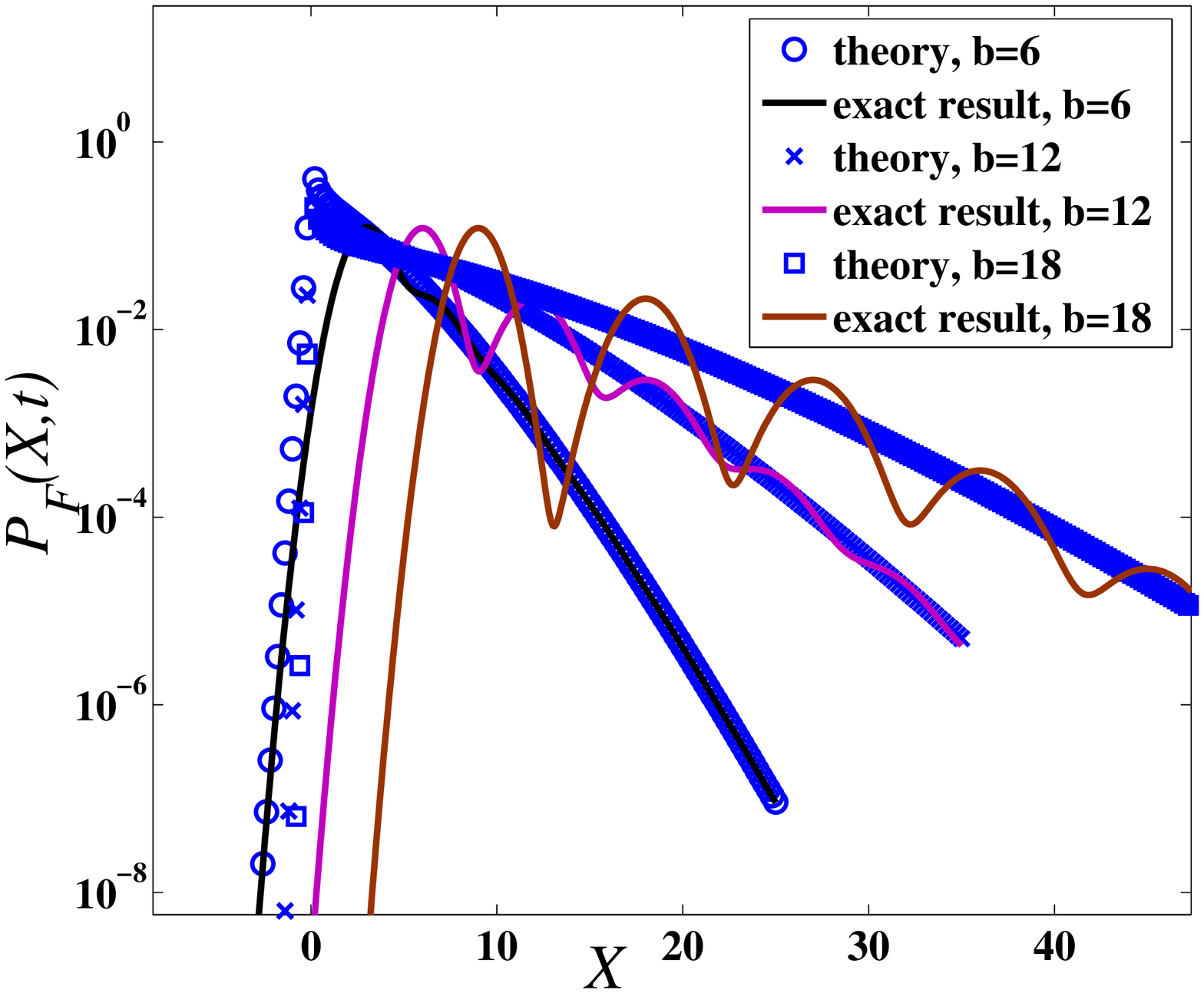}\\
 \caption{Presence of oscillation of the positional PDF in the limit of large $b$/strong forces. $\psi(\tau)=\exp(-\tau)$ and the displacements are drawn from Gaussian distribution (see Eq.~\eqref{GausssianDis}) with $\delta=1$. We choose $t=0.5$. The exact result are obtained from the first line of Eq.~\eqref{SMeq20301} and the theoretical predication Eq.~\eqref{SMeq2030111} with $P_0(X,t)$ calculated from Eq.~\eqref{SMeq20512}.}
 \label{ExpGaussianPXTStrong}
\end{figure}

\begin{figure}[tb]
  \centering
  \includegraphics[width=0.5\linewidth]{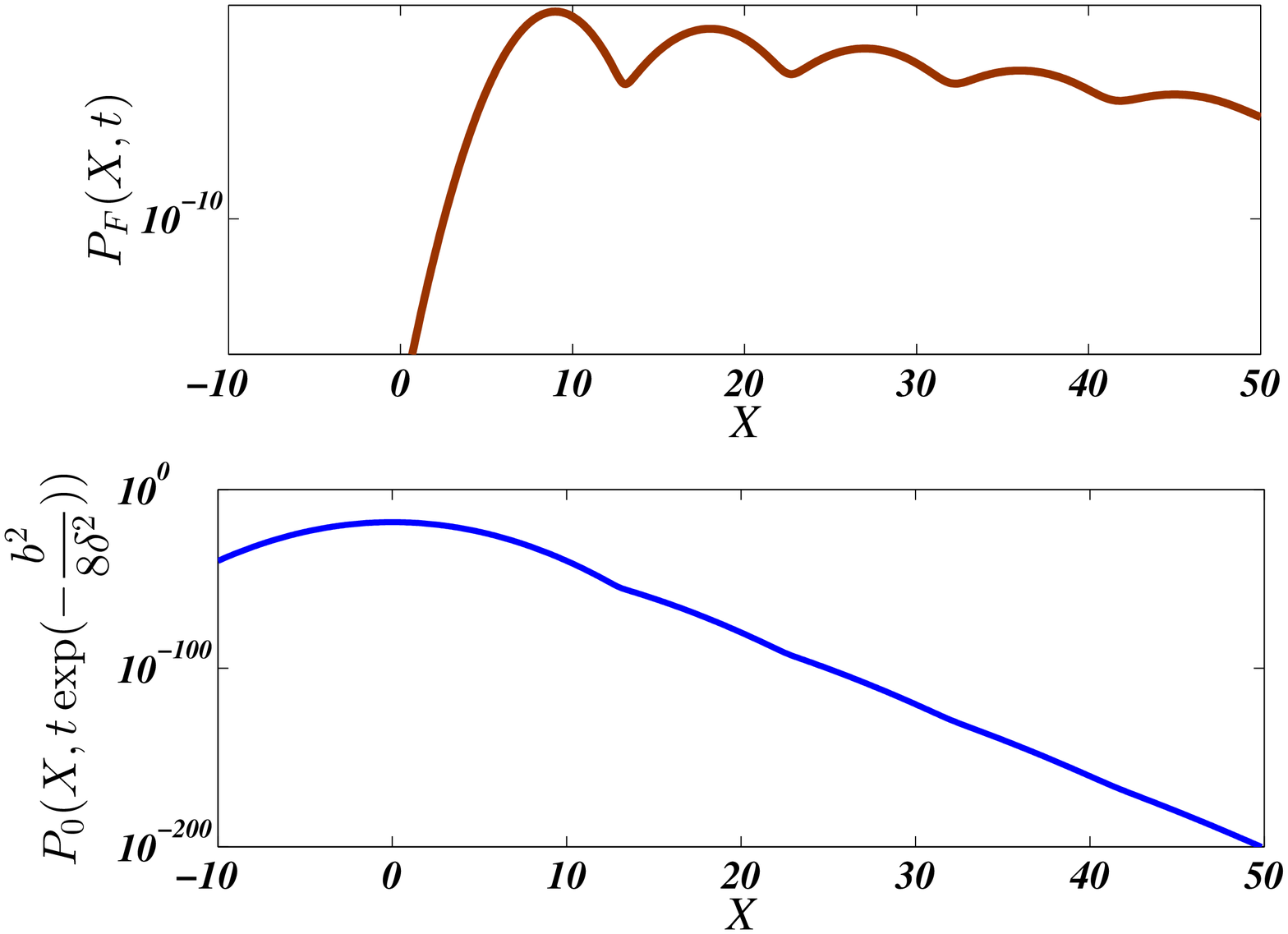}\\
  \caption{Plot of $P_F(X,t)$ [Eq.~\eqref{SMeq2030111}] and $P_0(X,t\exp(-\frac{b^2}{8\delta^2}))$ [Eq.~\eqref{SMeqnonbiase}] showing the effect of $\exp\left(\frac{bX}{2\delta^2}\right)\exp\left(-\left(1-\exp\left(-\frac{b^2}{8\delta^2}\right)\right)\frac{t}{\langle\tau\rangle}\right)$. $\psi(\tau)=\exp(-\tau)$ and the displacements are drawn from Gaussian distribution (see Eq.~\eqref{GausssianDis}) with $\delta=1$. Measurement time $t=0.5$ and $b=18$.   }
 \label{RelationBetweenBiasAndNon}
\end{figure}

\begin{figure}[tb]
\centering
\includegraphics[width=0.5\linewidth]{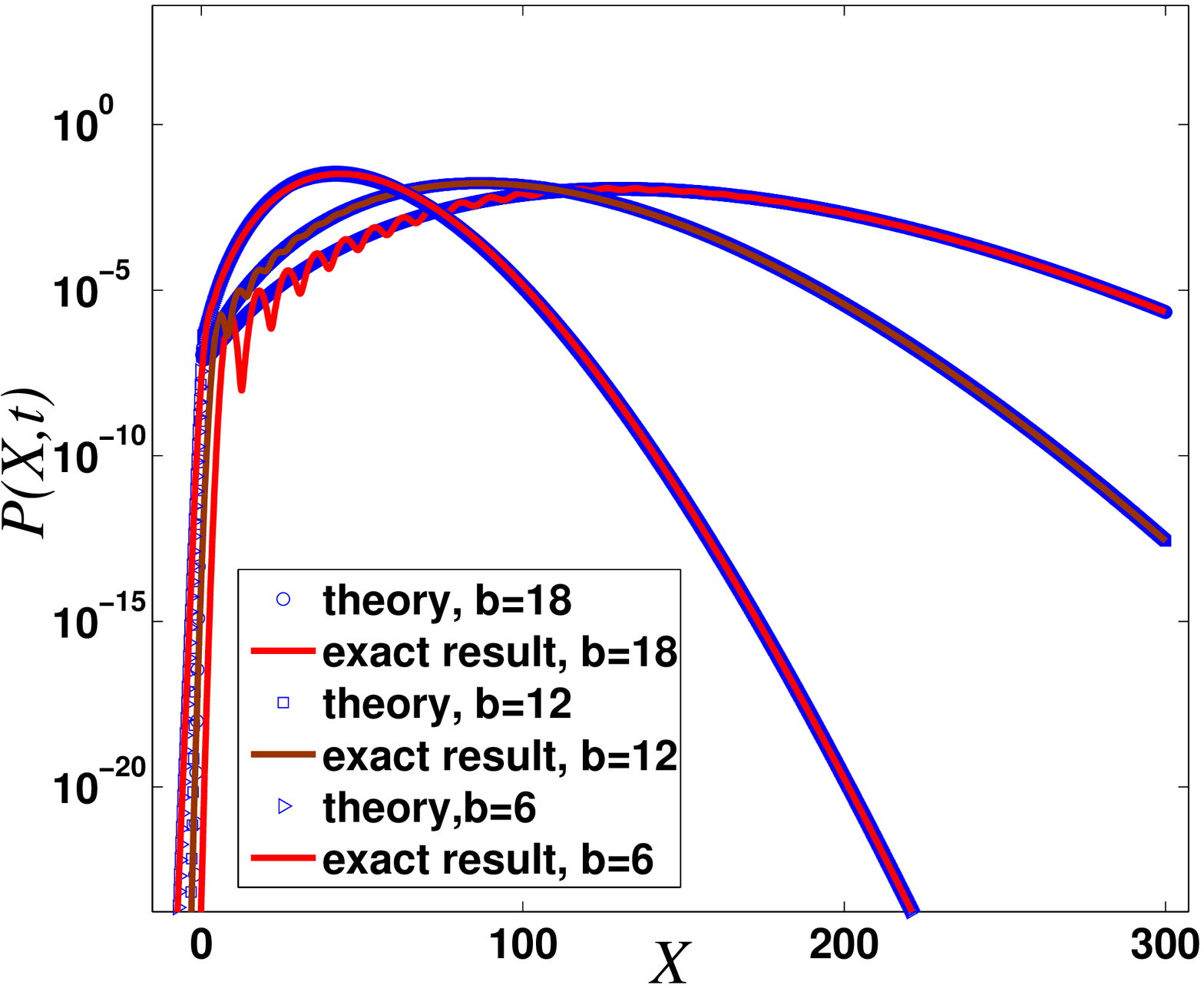}
\caption{Presence of oscillation of the positional PDF in the limit of large $b$/strong forces. $\psi(\tau)=\exp(-\tau)$ and the displacements are drawn from Gaussian distribution (see Eq.~\eqref{GausssianDis}) with $\delta=1$. We choose $t=15$. The exact result are obtained from the first line of Eq.~\eqref{SMeq20301} and the theoretical predication Eq.~\eqref{SMeq2030111} with $P_0(X,t)$ calculated from Eq.~\eqref{SMeq20512}. }\label{StrongBiasLongTime}
\end{figure}

\begin{figure}[htb]
  \centering
  \includegraphics[width=0.5\linewidth]{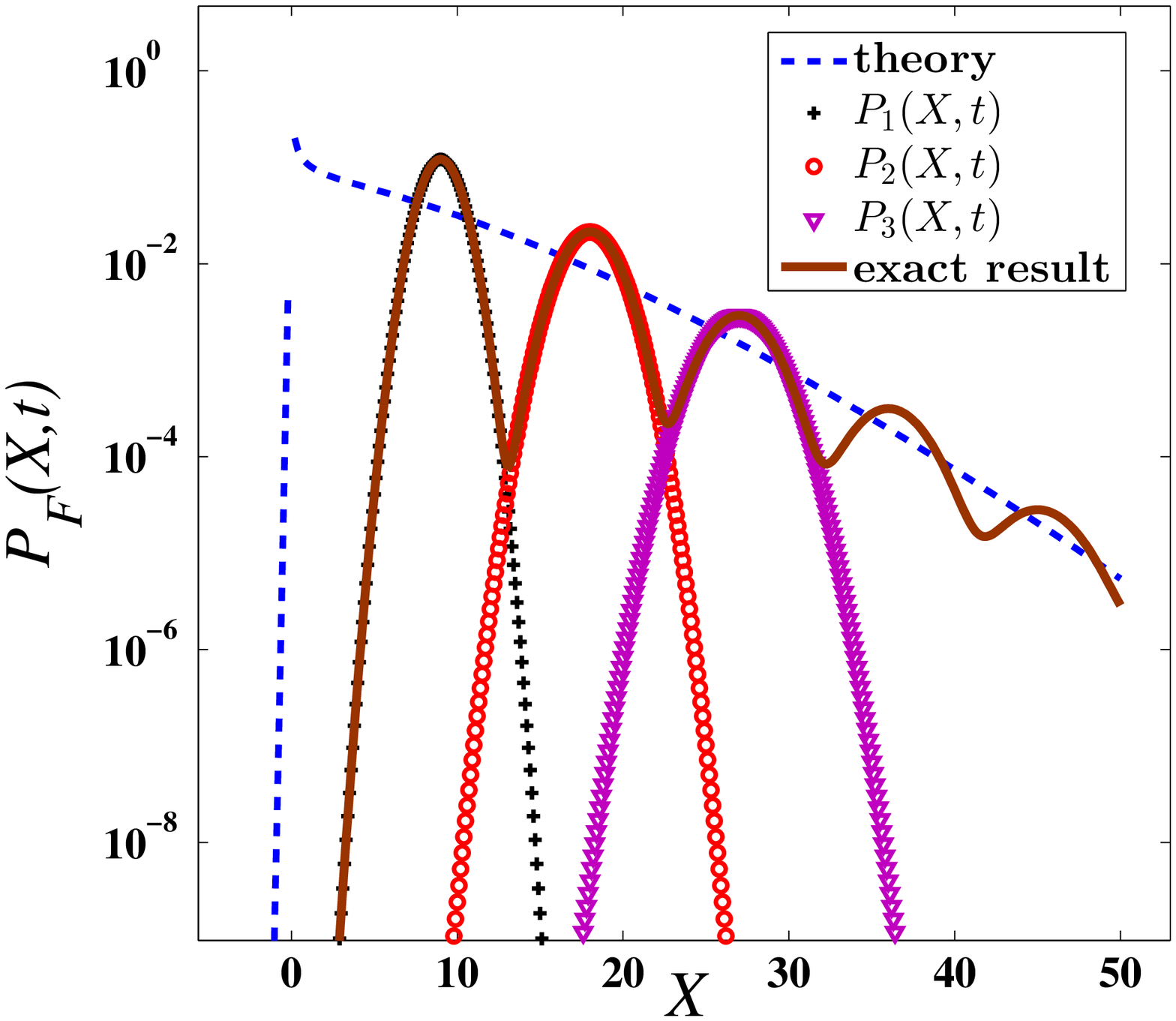}\\
  \caption{Tails of the positional distribution with a short observation time $t$ for a strong external force. $\psi(\tau)=\exp(-\tau)$ and the displacements are drawn from Gaussian distribution (see Eq.~\eqref{GausssianDis}) with $\delta=1$. The solid line is the exact result obtained from Eq.~\eqref{SMeq20301} and the dash line corresponds to Eqs.~\eqref{SMeq2030111} and \eqref{SMeq20512}. The symbols are related to $P_N(X,t)=Q_t(N)P_N(X)$ for a fixed $N$.
  Here $b=18$, and $t=0.5$. It can be seen that the multiple maxima of spreading packet are described by the fixed $N$, i.e., $P_N(X,t)$. }
 \label{ExpGaussianPXTStrongt05b18}
\end{figure}

We have shown that the relation between  an unperturbed system and a perturbed system, i.e., Eq.~\eqref{ewssj101}, is not limited to a weak bias/small $b$. But when we increase $b$ and approach the limit of strong forces, the behavior of $P_F(X,t)$ presents an interesting behavior. Oscillations of $P_F(X,t)$, instead of a monotonic decay, start to appear. See Fig.~\ref{ExpGaussianPXTStrong}.  
It may appear strange, since according to Eq.~\eqref{SMeq2030111} $P_F(X,t)$ is simply $P_0(X,\exp(-b^2/8\delta^2)t)$ multiplied by $\exp(b X/2\delta^2)\exp(-(1-\exp(-b^2/8\delta^2)t/\langle \tau \rangle)$. Because $P_0(X,t)$ decays monotonously we expect that multiplication of $P_0(X,t)$ by a monotonous function will produce a monotonous decay. But Fig.~\ref{RelationBetweenBiasAndNon} shows that $P_0(X,t)$ also attains small modulations. These modulations are extremely small and "magnified" by the huge prefactor $\exp(b X/2\delta^2)$ ($b$ is large) of $P_0(X,t)$. When $|X|$ is sufficiently large these oscillations disappear (see Fig.~\ref{StrongBiasLongTime}) and Eq.~\eqref{eq:unbiasedLong} holds. 
Fig.~\ref{ExpGaussianPXTStrongt05b18} shows that the oscillations of the positional PDF of driven system correspond to discrete events of fixed number of jumps. This means that for large $b$ and not sufficiently large $|X|$ the integral approximation for the sum (like the one in Eq.~\eqref{SMeq203}) fails. In such a case the probability to reach specific $|X|$ at time $t$ is dominated by very specific $N$.


\vspace{7cm}
\section{Universal exponential decay of the positional distribution}
Similar to the discussion in previous section, we discuss the general relation between $P_F(X,t)$ and $P_{0}(X,t)$, and use the tails of $P_{0}(X,t)$ to calculate  the large deviations of the positional distribution with bias.
\subsection{General relation between $P_F(X,t)$ and $P_{0}(X,t)$}
Here the aim is to extend Eq.~\eqref{SMeq2030111} considered in last section. For the waiting times, we follow the strategy given in Ref. \cite{BarkaiBurov2020} when $\psi(\tau)$ is analytic in the limit of $\tau\to 0$ , i.e., $$\psi(\tau)=C_A\tau^A+C_{A+1}\tau^{A+1}.$$
Here $A\geq 0$ is an integer. Based on \eqref{SMeq102}, the large $N/t$ limit leads to \cite{BarkaiBurov2020}
\begin{equation}\label{2sect01}
Q_t(N)\underset{N/t\to\infty}{\sim} \frac{\left(\left[C_A\Gamma(A+1)\right]^{\frac{1}{A+1}}t\right)^{N(A+1)}}{\Gamma\left[(A+1)N+1\right]}\exp\left(\frac{C_{A+1}}{C_A}t\right).
\end{equation}
For exponential distribution of waiting times with $\langle \tau \rangle =1$,  $\psi(\tau)\sim 1-\tau$, i.e., $A=0, C_A=1, C_{A+1}=-1$. From Eq.~\eqref{2sect01},  Eq.~\eqref{SMeq201} is obtained again.
For the displacement, we consider
\begin{equation}\label{disPhi}
\phi_{x\to x'}=f(x-x')\exp\left(\frac{F(x-x')}{2k_BT}\right)=f_0(x-x')\exp\left(-\lambda^2_F\right)\exp\left(\frac{F(x-x')}{2k_BT}\right),
\end{equation}
where $f_0(x)$ is the symmetric jump distribution of the unperturbed process.
As mentioned in the main text, $\lambda_F$ is a $F$-dependent and real constant. Note that when $F\to 0$ also $\lambda_F\to 0$.  Due to detailed balance condition (Eq. (4) in the main text) $f_0(x)$ must decay faster than exponential.

\begin{figure}[t]
 \centering
 \includegraphics[width=0.5\textwidth]{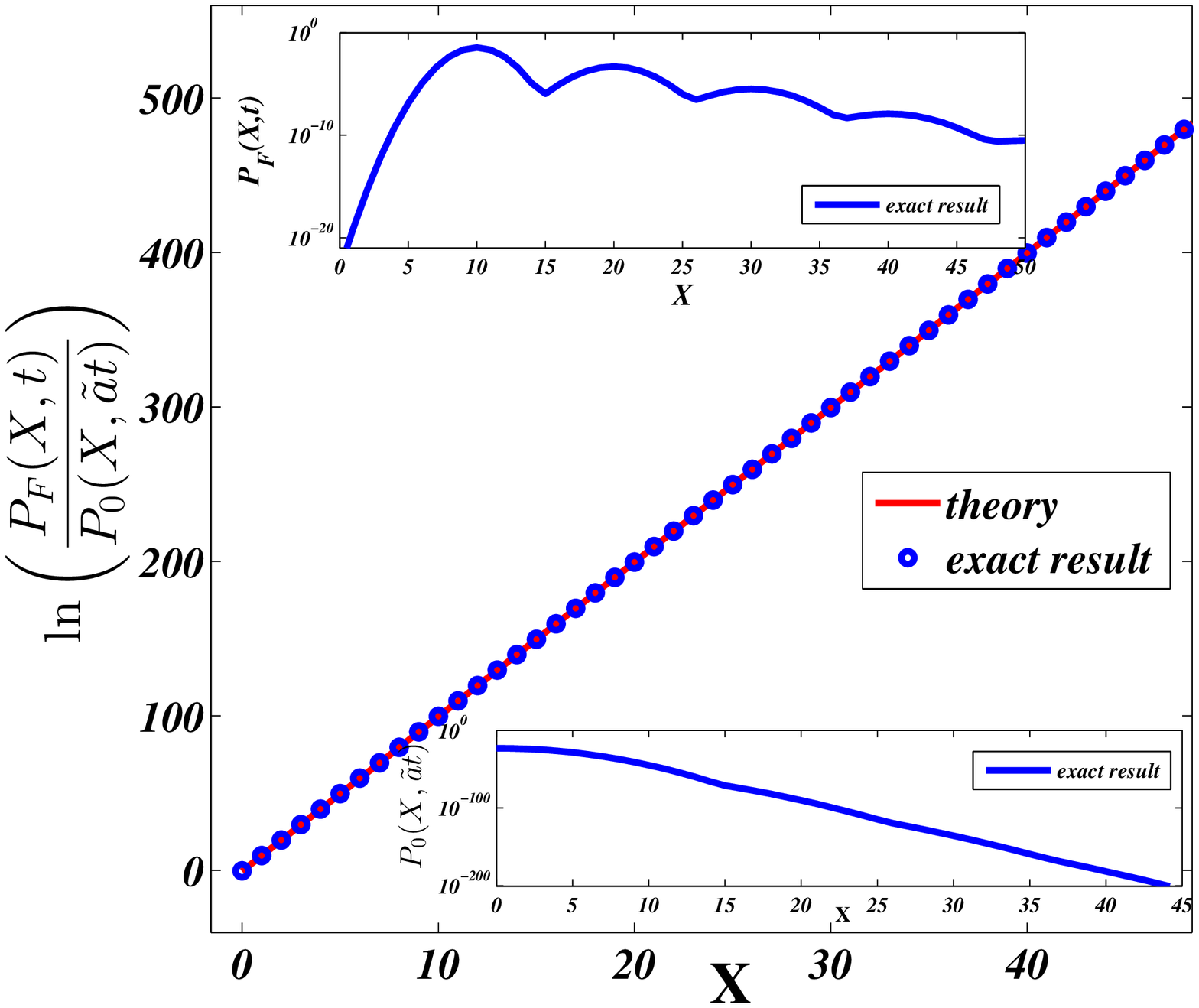}\\
 \caption{Plot of ${\rm ln}\left(\frac{P_F(X,t)}{P_0(X,\tilde{a}t)}\right)$, together with $P_F(X,t)$ and $P_0(X,\tilde{a}t)$. Here, we depict the correlation between $P_F(X,t)$ and $P_0(X,\tilde{a}t)$ predicted by Eq.~\eqref{2sect02addS101} and the corresponding exact results are obtained from Eq.~\eqref{SMeq101}. The waiting times are drawn from Erlang distribution $\psi(\tau)=\tau^{m-1}\exp(-\tau)/(m-1)!$ with $m=2$ and the displacements are generated from Eq.~\eqref{expressionfo} with $\delta=\sqrt{2}$, $\beta=2$, and $\lambda_F^2=50$, $F/(k_BT)=20$.
 }\label{PxtVsP0xt}
\end{figure}

The distribution of the position is provided by Eq.~\eqref{SMeq101} while ${\cal P}_N(X)$ is written as
\begin{equation}
{\cal P}_N(X) =
   \int\dots\int \phi_{0\to x_1}\dots \phi_{x_{N-1}\to X}\prod\limits_{i=1}^{N-1}dx_i.
    \label{eq:pnxmult}
\end{equation}
Eq.~\eqref{eq:pnxmult} together with Eq.~\eqref{disPhi} yields
\begin{equation}
{\cal P}_N(X) = \exp\left(\frac{F}{2k_BT}X-\lambda^2_FN\right)P_{0,N}(X),
    \label{eq:pnxp0x}
\end{equation}
while $P_{0,N}(X)$ is the PDF function of $X$ given that exactly $N$ transition were performed under the condition of $F=0$. Therefore $P_F(X,t)$ attains the form
\begin{equation}
P_F(X,t)=
\exp\left(\frac{F}{2k_BT}X\right)
\sum_{N=0}^\infty \exp\left(-\lambda^2_FN\right)P_{0,N}(X)Q_t(N)
\label{eq:pxtgenformS}
\end{equation}
The limit of large $X$ is determined by the limit of large $N$ (see below). Then we can use the form of $Q_t(N)$ provided in Eq.~\eqref{2sect01} and write
\begin{equation}
\exp\left(-\lambda^2_FN\right)Q_t(N) = \exp\left(\frac{C_{A+1}}{C_A}[1-\tilde{a}]t\right) 
Q_{{\tilde a}t}(N),
    \label{eq:qtntransform}
\end{equation}
where 
\begin{equation}
\tilde{a}=\exp(-\lambda_F^2/(A+1)).
    \label{eq:atilde}
\end{equation}
Then Eq.~\eqref{eq:pxtgenformS} yields
\begin{equation}\label{2sect02addS101}
P_F(X,t)\underset{|X|\to\infty}{\sim} \exp\left(\frac{F}{2k_BT}X+\frac{C_{A+1}}{C_A}[1-\tilde{a}]t\right)\sum_{N=0}^\infty Q_{\tilde{a}t}(N)P_{0,N}(X),
\end{equation}
and finally we obtain
\begin{equation}\label{2sect02addS101addf}
P_F(X,t)\underset{|X|\to\infty}{\sim} \exp\left(\frac{F}{2k_BT}X+\frac{C_{A+1}}{C_A}[1-\tilde{a}]t\right)P_0(X, \tilde{a}t),
\end{equation}
where $P_0(X, \tilde{a}t)=\sum_{N=0}^\infty Q_{\tilde{a}t}(N)P_{0,N}(X)$ is the probability of the unperturbed process to occupy $X$ at time ${\tilde a}t$. In Fig.~\ref{PxtVsP0xt} we display a perfect match between the result in Eq.~\eqref{2sect02addS101addf} and the exact behavior for the specific case of Erlang distribution of waiting times and Gaussian $f_0(x)$. 
 Below we  investigate the far tails of $P_F(X,t)$, which according to Eq.~\eqref{2sect02addS101addf} reduces to the problem of the far tails of $P_0(X, t)$.

\subsection{Decay of the positional distribution with a general waiting time distribution}
Based on Eq.~\eqref{disPhi}, for $f_0(x)$ we focus on
\begin{equation}\label{expressionfo}
f_0(x)=c\exp\left(-\left(\frac{|x|}{\delta}\right)^\beta\right),
\end{equation}
where $c$ is the normalizing constant. As previously mentioned, we assume that $f_0(x)$ decays faster than
exponential, i.e., $\beta\geq 1$.

First we write the form of $P_{0,N}(X)$ for the case of  Eq.~\eqref{expressionfo} and then use it to estimate the tails of $P_0(X,t)$.
From Eq.\eqref{expressionfo} and in Fourier space we have
\begin{equation}\label{2sect02}
\begin{split}
\widetilde{f}_0(k)&=\int_{-\infty}^\infty \exp(ikx)f_0(x)dx\\
&\sim \frac{c\sqrt{2\pi}}{\sqrt{\beta^{\frac{1}{\beta-1}}(\beta-1)\delta^{-\frac{\beta}{\beta-1}}}}\exp\left(\frac{(\beta-1)\delta^{\frac{\beta}{\beta-1}}}{\beta^{\frac{\beta}{\beta-1}}}|ik|^{\frac{\beta}{\beta-1}}\right),
\end{split}
\end{equation}
while $x$ is treated as a large variable. Note that the displacements of the particles are IID random variables. Then $P_{0,N}(x)$ is provided by
\begin{equation}\label{2sect03}
\begin{split}
P_{0,N}(X)&=\frac{1}{2\pi}\int_{-\infty}^\infty \exp(-ikX) (\widetilde{f}_0(k))^Ndk\\
&\sim \frac{(c\sqrt{2\pi})^N}{\sqrt{\beta^{\frac{1}{\beta-1}}(\beta-1)\delta^{-\frac{\beta}{\beta-1}}}} \frac{\exp(-N\frac{|X|^\beta}{(N\delta)^\beta})}{\sqrt{2\pi N^{\beta-1}\frac{\delta^\beta}{\beta(\beta-1)}|X|^{2-\beta}}}.
\end{split}
\end{equation}
The positional distribution without the bias is
\begin{equation}\label{2sect04}
\begin{split}
P_0(X,t)&=\sum_{N=0}^\infty Q_{t}(N)P_{0,N}(X)\sim\int_0^\infty\exp\Big(\Phi(X,t,N)\Big)dN,
\end{split}
\end{equation}
 and
\begin{equation}\label{PhiXtN}
\begin{split}
\Phi(X,t,N)&=-\ln \left(2 \pi  (\delta N)^{\beta/2} \sqrt{\frac{(A+1) | X| ^{2-\beta}}{\beta (\beta-1) }}\right)+(A+1) N \ln \left(\frac{e t (C_A \Gamma (A+1))^{\frac{1}{A+1}}}{(A+1) N}\right)\\
&~~~~~~+N \ln \left(\frac{\sqrt{2 \pi } c}{\sqrt{ \beta^{\frac{1}{\beta-1}} (\beta-1)  \delta^{-\frac{\beta}{\beta-1}}}}\right)-N \left(\frac{| X| }{\delta N}\right)^{\beta}+\frac{C_{A+1} }{C_A}t.
\end{split}
\end{equation}
\begin{figure}[t]
 \centering
 \includegraphics[width=0.5\textwidth]{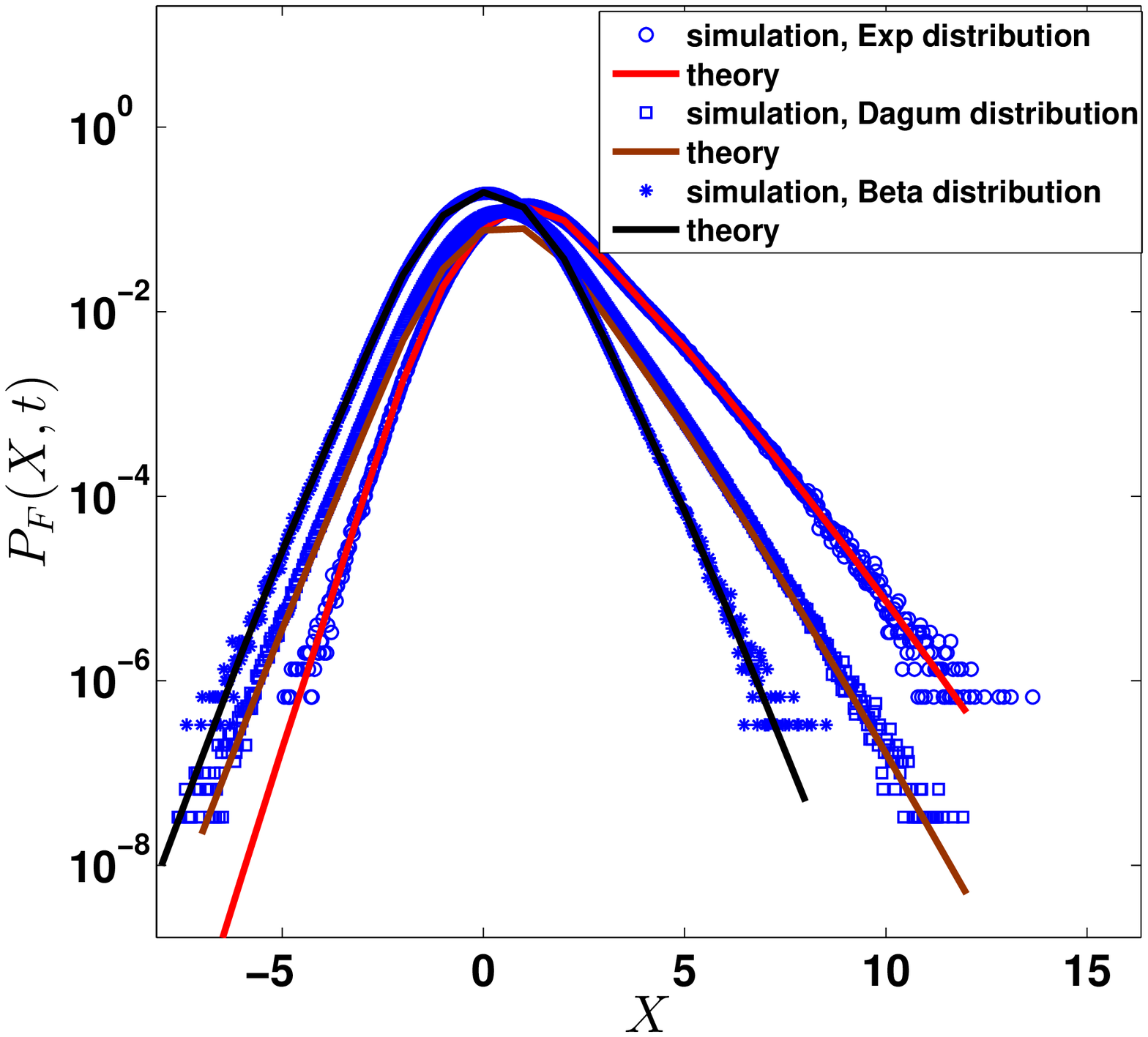}\\
 \caption{Asymmetric exponential tails of the positional distribution for different biases with various waiting time PDFs, i.e., exponential distribution $\psi(\tau)=\exp(-\tau)$ (`$\square$'), Dagum distribution $\psi(\tau)=1/(1+\tau)^2$ (`$\circ$') and a special form of Beta distribution $\psi(\tau)=6\tau(1-\tau)$ (`$\ast$'). The corresponding displacement are drawn from Eq.~\eqref{disPhi} and $f_0(x)=1/\sqrt{2\pi}\exp(-|x/\delta|^2)$ with $\lambda^2_F=1/2$, $F/K_BT=2$; and $\lambda^2_F=1/8$, $F/K_BT=1$; $\lambda^2_F=1/200$, $F/K_BT=1/5$, respectively. The symbols are simulations obtained from $10^9$ realizations and the solid lines are theoretical predictions Eq.~\eqref{2sect02addS101addf} without considering the non-moving particles, where $Q_t(N)$ and $P_{0,N}(X)$ are obtained from \eqref{2sect01} and Eq.~\eqref{2sect03}. 
}\label{fig:pxttailsSM}
\end{figure}

 We exploit the saddle point approximation in order to solve the integral in Eq.~\eqref{2sect04}, and assume that $|X|\to\infty$, therefore  
\begin{equation}\label{PzeroForm}
P_0(X,t)\sim \sqrt{2\pi}\frac{\exp(\Phi(X,t,N^*))}{\sqrt{|\Phi^{''}(X,t,N^*)|}},
\end{equation}
where $N^*$ obeys
\begin{equation}\label{Nxing}
\frac{\partial}{\partial N} \Phi(X,t,N)|_{N=N^*}=0
\end{equation}
and the second-order derivative
\begin{equation}
\Phi^{''}(X,t,N^*)\sim -(\beta-1) \beta \delta^{-\beta} (N^*)^{-\beta-1} | X| ^{\beta}.
\end{equation}
Based on Eq.~\eqref{PzeroForm}, for large $|X|$ we obtain
\begin{equation}\label{PxtNxing}
P_0(X,t)\sim \frac{\exp\left(N^{*} \ln \left(\frac{\sqrt{2 \pi } c (et)^{A+1} C_A \Gamma (A+1)\delta^{\frac{\beta}{2 (\beta-1)}}}{\beta^{\frac{1}{2 (\beta-1)}} \sqrt{\beta-1} ((A+1) N^{*})^{A+1} }\right)-N^{*} \left(\frac{| X| }{\delta N^{*}}\right)^{\beta}+\frac{C_{A+1} }{C_A}t\right)}{\sqrt{A+1} \sqrt{\frac{2 \pi }{N^{*}}} | X| }.
\end{equation}
Next step is to find the asymptotic solution of Eq.~\eqref{Nxing}, which yields
\begin{equation}\label{nxing}
N^{*}\sim \frac{|X|}{\delta}\left(\frac{(\beta-1) \beta /(A+1)}{ W_0\left(\frac{(\beta-1) \beta }{A+1}(2 \pi )^{-\frac{\beta}{2 A+2}} \left(\frac{1}{\delta}\right)^{\beta} \left| (\beta-1) \beta^{\frac{1}{\beta-1}}\right| ^{\frac{\beta}{2 A+2}} | X| ^{\beta} \left(\frac{\sqrt{\delta^{-\frac{\beta}{\beta-1}}} \left(\frac{t (C_{A} \Gamma (A+1))^{\frac{1}{A+1}}}{A+1}\right)^{-A-1}}{c}\right)^{\frac{\beta}{A+1}}\right)}\right)^{1/\beta}.
\end{equation}
When $|X|\to +\infty$, the asymptotic behavior between $N^*$ and $X$ is 
\begin{equation}\label{asyNxing}
N^{*}\propto 
\frac{|X|/\delta}{\ln\left(|X|\Big/C_A^{\frac{1}{A+1}}\delta t\right)},
\end{equation}
demonstrating a nearly linear scaling of $N^*$ with $X$. 

 Plugging Eq.~\eqref{nxing} into Eq.~\eqref{PzeroForm}, the far tails of the positional distribution obey
\begin{equation}\label{nonbiased}
P_0(X,t)\underset{|X|\to\infty}{\sim} \frac{\exp\left(\frac{C_{A+1}}{C_A}t-|X|\frac{g^{\frac{1}{\beta}}}{(W_0(h|X|^\beta/ t^\beta))^{\frac{1}{\beta}}}\left(\frac{W_0(h(\frac{|X|}{t})^\beta)}{g\sigma^\beta}-\ln(M (\frac{t}{|X|})^{A+1}(W_0(g(\frac{|X|}{t})^\beta))^{\frac{A+1}{\beta}})\right)\right)}{\sqrt{\frac{2\pi\delta(A+1)^{1+\frac{1}{\beta}}W_0^{\frac{1}{\beta}}(h|X|^\beta/t^\beta)|X|}{(\beta(\beta-1))^{\frac{1}{\beta}}}}},
\end{equation}
where
\begin{equation}
g=\frac{(\beta-1)\beta}{\sigma^\beta(A+1)},  
\end{equation}
\begin{equation}
h=g\sqrt[A+1]{\frac{((\beta-1)^{\frac{1}{\beta-1}}\beta)^{\frac{\beta}{2}}}{c(2\pi)^{\frac{\beta}{2}}[C_A\Gamma[A+1]]^\beta\delta^{\frac{\beta^2}{2(\beta-1)}}}},
\end{equation}
and
\begin{equation}
M=\frac{\sqrt{2 \pi } c C_A \exp (A+1) \Gamma (A+1) \delta^{A+\frac{\beta}{2 (\beta-1)}+1}}{\beta^{\frac{A+1}{\beta}+\frac{1}{2 (\beta-1)}} (\beta-1)^{\frac{A+1}{\beta}+\frac{1}{2}} (A+1)^{\frac{(A+1) (\beta-1)}{\beta}}}.
\end{equation}
See also the calculation and related discussion in Ref.\cite{BarkaiBurov2020}.

Equation~\eqref{nonbiased} together with Eq.~\eqref{2sect02addS101addf} prove that the decay of of positional PDF of biased process, i.e., $P_F(X,t)$, is exponential (up to logarithmic corrections) and asymmetric (see Fig.~\ref{fig:pxttailsSM}).

\end{document}